\documentclass[journal=jpcafh,manuscript=article]{achemso}
\usepackage{longtable}
\usepackage{multirow}
\usepackage{amsmath}
\usepackage{subfigure}
\usepackage[usenames,dvipsnames]{color}
\usepackage{soul}
\usepackage{bm}

\usepackage{xr}
\usepackage{amssymb}

\usepackage{booktabs}

\usepackage{multicol} \usepackage{wrapfig} \usepackage{enumitem}
\usepackage{bm} \usepackage{outlines} 

\usepackage{multicol} \usepackage{wrapfig} \usepackage{enumitem}
\SectionNumbersOn

\author{Silvan K\"aser} \affiliation[University of Basel]{Department
  of Chemistry, University of Basel, Klingelbergstrasse 80 , CH-4056
  Basel, Switzerland.}

\author{Eric Boittier} \affiliation[University of Basel]{Department
of Chemistry, University of Basel, Klingelbergstrasse 80 , CH-4056
Basel, Switzerland.}

\author{Meenu Upadhyay} \affiliation[University of Basel]{Department
of Chemistry, University of Basel, Klingelbergstrasse 80 , CH-4056
Basel, Switzerland.}

\author{Markus Meuwly} \affiliation[University of Basel]{Department of
  Chemistry, University of Basel, Klingelbergstrasse 80 , CH-4056
  Basel, Switzerland.}  \email{m.meuwly@unibas.ch}

\title{MP2 Is Not Good Enough: Transfer Learning ML Models for Accurate
  VPT2 Frequencies}
\begin{document}

\date{\today}
\section*{Abstract}
\label{sec:Abstract}
The calculation of the anharmonic modes of small to medium sized
molecules for assigning experimentally measured frequencies to the
corresponding type of molecular motions is computationally challenging
at sufficiently high levels of quantum chemical theory. Here, a
practical and affordable way to calculate coupled-cluster quality
anharmonic frequencies using second order vibrational perturbation
theory (VPT2) from machine-learned models is presented. The approach,
referred to as ``NN + VPT2'', uses a high-dimensional neural network
(PhysNet) to learn potential energy surfaces (PESs) at different
levels of theory from which harmonic and VPT2 frequencies can be
efficiently determined. The NN + VPT2 approach is applied to eight
small to medium sized molecules (H$_2$CO, trans-HONO, HCOOH, CH$_3$OH,
CH$_3$CHO, CH$_3$NO$_2$, CH$_3$COOH and CH$_3$CONH$_2$) and
frequencies are reported from NN-learned models at the
MP2/aug-cc-pVTZ, CCSD(T)/aug-cc-pVTZ and CCSD(T)-F12/aug-cc-pVTZ-F12
levels of theory. For the largest molecules and at the highest levels
of theory, transfer learning (TL) is used to determine the necessary
full-dimensional, near-equilibrium PESs. Overall, NN+VPT2 yields
anharmonic frequencies to within 20 cm$^{-1}$ of experimentally
determined frequencies for close to 90 \% of the modes for the highest
quality PES available and to within 10 cm$^{-1}$ for more than 60 \%
of the modes. For the MP2 PESs only $\sim 60$~\% of the NN+VPT2
frequencies were within 20~cm$^{-1}$ of the experiment, with outliers
up to $\sim 150$~cm$^{-1}$ compared with experiment.  It is also
demonstrated that the approach allows to provide correct assignments
for strongly interacting modes such as the OH bending and the OH
torsional modes in formic acid monomer and the CO-stretch and OH-bend
mode in acetic acid.

\section{Introduction}
\label{sec:Introduction}
Vibrational spectroscopy is a sensitive probe for identifying
molecules or to follow conformational and structural changes in the
gas phase and in solution. One essential task in the practical use of
vibrational spectroscopy is the assignment of a measured frequency to
its corresponding type of molecular motion. Based on this information
it is also possible to predict changes both in the characterization of
these motions and their influence on the frequencies themselves.  In
spectrally congested regions, as in the frequency range between 1200
cm$^{-1}$ and 1700 cm$^{-1}$, these assignments are particularly
challenging due to couplings between the different degrees of freedom.
Similarly, force field parametrization relies on fitting computed
normal mode frequencies to the correctly assigned band positions from
experiment. In practice it would, however, be preferable to use
anharmonic computed frequencies in force field development because
normal modes are already based on the harmonic oscillator
assumption. It is for such tasks that computational approaches are
particularly valuable.  \\

\noindent
An accurate description of the vibrational dynamics and IR
spectroscopy remains a challenging problem in molecular
spectroscopy\cite{qu2019quantum}.  Often, these calculations require
accurate, full-dimensional potential energy surfaces (PESs) for which
machine learning (ML) methods have gained a lot of
attention\cite{unke2020high}. ML potentials are used to generate
statistical models for energies and forces based on molecular
structures from extensive \textit{ab initio} data. The resulting
potentials can reproduce the reference data with unprecedented
accuracies\cite{meuwly2020ml,qu2021acac} (energies and forces with
errors in the range of $10^{-2} - 10^{-5}$~kcal/mol and $10^{-1} -
10^{-3}$~kcal/mol/\r{A}, harmonic frequencies are obtained within
$\sim 0.5$~cm$^{-1}$) and thus are superior to \textit{ab initio}
potentials due to their efficiency.\\

\noindent
Predictions of accurate anharmonic frequencies which compare
sufficiently well with experiment remain a challenge to overcome and
allow assignments of the vibrations and interpretation of
spectroscopic
features.\cite{nejad2020glycolic,mata2017benchmarking,barone2014fully}
Recent studies\cite{qu2018ir,qu2018quantum} illustrate that IR spectra
determined from molecular dynamics (MD) simulations are unable to
capture the full anharmonic behaviour of the molecule, especially for
high-frequency modes. This motivates the search for alternative
approaches, given that a sufficient number of reference calculations,
at high levels of theory, have become
possible\cite{qu2018high,koner2020permutationally,meuwly2020ml}.\\

\noindent
This work presents a practical approach to calculate anharmonic
frequencies from accurate, machine-learned potentials for small (4
atoms) to medium-sized (9 atoms) molecules. The ML-PESs are used as an
external energy function to quantum chemistry software (here the
Gaussian package\cite{g09}), to determine energies, forces, force
constants, dipole moments and dipole moment derivatives for a
vibrational perturbation theory (VPT2)
analysis\cite{barone2005anharmonic}. This has the advantage that a
direct comparison of results from explicit normal mode and VPT2
calculations, at a given level of theory from the electronic structure
code, and the results from using the ML-PES is possible. Moreover, ML
models can be systematically improved by including additional data
and/or by using data from higher levels of quantum chemical
theory. For this, transfer learning (TL) schemes, which have been
shown to be data-efficient alternatives
\cite{taylor2009transfer,pan2009survey,meuwly2020ml}, can be used to
achieve higher quality PESs and will be explored here, too. \\

\noindent
The present work reports on a systematic study of anharmonic
frequencies for H$_2$CO, trans-HONO, HCOOH, CH$_3$OH, CH$_3$CHO,
CH$_3$NO$_2$, CH$_3$COOH and CH$_3$CONH$_2$ based on machine-learned
models employing reference data from electronic structure calculations
at different levels of theory. The machine-learned models will then be
used for harmonic and VPT2 calculations which can also be compared
with results from experiments. Furthermore, TL to higher levels of
theory is explored. As an example, for the largest molecules a
sufficient number of energies and forces (estimated to be around
$10^4$ or larger) at the highest levels of theory, such as CCSD(T), is
usually not feasible. Hence, a relevant question is whether by
starting from a robust MP2-learned PES one can `transfer learn' to a
CCSD(T)-quality PES from a considerably smaller set of reference
energies and gradients at this higher level of theory which also
yields improved anharmonic frequencies compared with the MP2
level. Earlier work suggests that TL can result in substantial
improvements in accuracy and
data-efficiency\cite{meuwly2020ml}. Finally, the study will also
discuss how the cost for generating an accurate ML model (i.e. the
cost of structure sampling, training and evaluating the ML model)
compares with a single, 'straight' \textit{ab initio} VPT2
calculation. \\

\noindent
First, the methods and data generation strategies are presented. This
is followed by an assessment of the harmonic and anharmonic modes from
the ML-PESs compared with reference electronic structure calculations
at different levels of theory. Also, where available, comparison with
experimental results is made. Then, the computational efficiency of
the approach chosen is considered. Finally, 
the results are discussed in a broader context and conclusions are drawn.

\section{Computational Methods}
\label{sec:Methods}

\subsection{Data sets: sampling and quantum chemical methods}
Data sets at three levels of theory, including
MP2/aug-cc-pVTZ\cite{moller1934note, kendall1992electron},
CCSD(T)/aug-cc-pVTZ\cite{pople1987quadratic, purvis1982full,
  kendall1992electron}, and
CCSD(T)-F12/aug-cc-pVTZ-F12\cite{adler2007simple,
  peterson2008systematically} (referred to as ``MP2'', ``CCSD(T)'',
and ``CCSD(T)-F12'' for convenience in the following), were
generated. All single point electronic structure calculations,
including energies, forces and dipole moments required for ML, as well
as harmonic frequency calculations, were carried out using
MOLPRO\cite{MOLPRO}. Data sets at the MP2 level of theory were
generated for all molecules, for CCSD(T) they were generated for
molecules with $N_{\rm atom} \leq 6$, and data sets at the CCSD(T)-F12
level of theory were generated for molecules with $N_{\rm atom} \leq
5$.\\

\noindent
As molecules of different sizes are considered, the number of
\textit{ab initio} calculations for the data base was scaled
accordingly. Here, $(3N-6)\cdot 600$ geometries were sampled for each
molecule and the optimized geometry was added. Generation of the
reference geometries was based on normal mode
sampling\cite{smith2017ani}: First, the molecules were optimized at
the MP2 level of theory, and the normal mode vectors were determined
alongside with the corresponding harmonic force constants. Then,
distorted (non-equilibrium) conformations were obtained by randomly
displacing the atoms along linear combinations of normal mode vectors.
This sampling was carried out at different temperatures (here $T =10$,
50, 100, 300, 500, 1000, 1500 and 2000~K). The total number of
geometries was evenly divided between the temperatures ,
i.e. $((3N-6)\cdot 600)/8$ geometries were generated for each $T$.\\

\noindent
For three molecules, HONO, CH$_3$CHO and CH$_3$COOH, the initial data
set was enlarged by including additional geometries as the PhysNet
models predicted normal mode frequencies with larger errors than for
the other molecules. For HONO, an additional 2805 structures were
added. The geometries were sampled from $NVT$ simulations run at
1000~K using the semiempirical GFN2-xTB method\cite{bannwarth2019gfn2}
and extended with structures along particular normal modes. Moreover,
because the structures evaluated in the VPT2 calculations (i.e. the
atoms are slightly displaced from the equilibrium geometry for the
calculation of numerical derivatives) can be extracted, these were
added as well (henceforth, ``VPT2 geometries''). For CH$_3$CHO, a
total of 1072 additional structures were added. These correspond to
VPT2 geometries and geometries along particular normal modes.  For
CH$_3$COOH, 109 additional, VPT2 geometries are added.\\

\subsection{Machine learning: PhysNet}
The representation of the PESs is based on a neural network (NN) using
the PhysNet architecture.\cite{MM.physnet:2019} A detailed description
of the NN architecture is given in Reference~\citenum{MM.physnet:2019}
and only the salient features and those used in the present work are
given below. PhysNet is a high-dimensional NN of the ``message
passing'' type\cite{gilmer2017neural} and was applied recently to
different molecular
systems\cite{rivero2019reactive,brickel2019reactive,mm.ht:2020,mm.atmos:2020,sweeny2020thermal,meuwly2020ml}. The
NN learns a feature vector that encodes the local chemical environment
of each atom $i$ for the prediction of total molecular energies,
atomic forces and molecular dipole moments. The feature vector
initially contains information about the nuclear charges $Z_i$ and Cartesian
coordinates $\bm{r}_i$ and is iteratively refined (``learned'') during
training. The total energy of a molecule with arbitrary geometry
includes long-range electrostatics and dispersion interactions and is
given by
\begin{align}
    E = \sum_{i=1}^N E_i +
    k_e\sum_{i=1}^N\sum_{j>i}\frac{q_iq_j}{r_{ij}} + E_{\rm D3}.
\end{align}
where $E_i$ are the atomic energy contributions, $E_{\rm D3}$ is
Grimme's D3 dispersion correction \cite{grimme2010consistent}, $q_i$
are partial charges, $r_{ij}$ are interatomic distances, $k_e$ is
Coulomb's constant and $N$ is the total number of atoms. Note that the
partial charges are adjusted to assure charge conservation and the
Coulomb term is damped for short distances to avoid numerical
instabilities, see Ref.~\citenum{MM.physnet:2019} for details. Besides
the energy, PhysNet also predicts molecular dipole moments from
partial charges according to $\bm{\mu} = \sum_{i=1}^N q_i \bm{r_i}$
and analytical derivatives of $E$ with respect to the Cartesian
coordinates of the atoms are obtained by reverse mode automatic
differentiation\cite{baydin2017automatic}.\\

\noindent
For the present work, PhysNet was adapted to additionally predict
analytical derivatives of the dipole moment $\bm{\mu}$ and second
order derivatives of $E$ with respect to Cartesian coordinates
(i.e. Hessians). This was easily achieved using
Tensorflow\cite{tensorflow2015-whitepaper} utilities.\\

\noindent
The data sets containing $(3N-6)\cdot 600 + 1$ data points were split
according to 85/10/5~\% into training/validation/test sets for the
fitting of the ML model. The training set always contained the
optimized geometry. PES representations are obtained by adapting the
PhysNet parameters to best describe reference energies, forces, and
dipole moments from explicit quantum chemical calculations.  The
optimization is carried out using
AMSGrad\cite{reddi2019convergence}. The relative contribution of the
different error terms is controlled by weighting hyperparameters
giving the force errror a higher weight/importance compared to the
energy error. All hyperparameters of the NN architecture and its
optimization approach were set to the values presented in
Reference~\citenum{MM.physnet:2019}, except for the cutoff radius
$r_{\rm cut}$ for interactions in the NN which was set to 6~\AA.  Such
a cutoff was sufficient to include all interactions in all
molecules.\\

\subsection{Vibrational second-order perturbation theory}
Second-order vibrational perturbation theory (VPT2) is used to include
anharmonic and mode coupling effects into spectroscopic
properties\cite{nielsen1951vibration}. Many of the common quantum
chemistry packages, such Gaussian\cite{g09,barone2005anharmonic} or
CFOUR\cite{cfour}, include implementations for the calculation of
infrared frequencies and intensities including anharmonic effects
directly from \textit{ab initio} data. VPT2 assumes that the potential
energy of a system can be expressed as a quartic force field given by
\begin{align}\label{eq:vpt2_forcefield}
    V = \frac{1}{2}\sum\omega_i\hat{q_i}^2 + \frac{1}{3!}\sum \phi_{ijk} \hat{q_i}
    \hat{q_j} \hat{q_k} + \frac{1}{4!} \sum\phi_{ijkl} \hat{q_i} \hat{q_j} \hat{q_k} \hat{q_l}.
\end{align}

\noindent
Here, $\omega_i$ is a harmonic frequency, $\hat{q_i}$ are (reduced
dimensionless) normal mode coordinates, not to be confused with the
partial charges $q_i$, and $\phi_{ijk}$ and $\phi_{ijkl}$ are third-
and fourth-order derivatives of the potential $V$ with respect to
normal mode coordinates
\cite{yu2015vibrational,barone2005anharmonic}. The first term in
Equation~\ref{eq:vpt2_forcefield} corresponds to the harmonic part of
the potential while the remaining terms describe anharmonic
effects. Using expressions from earlier
work\cite{yu2015vibrational,barone2005anharmonic,bloino2012general,bloino2015vpt2}
and omitting kinetic/rotational terms, the cubic and quartic force
constants can be used to obtain anharmonic constants:
\begin{align}
    16\chi_{ii} &= \phi_{iiii} - \sum_j
    \frac{\left(8\omega_i^2-3\omega_j^2\right)\phi^2_{iij}}{\omega_j\left(4\omega_i^2
      - \omega_j^2\right)}\\
   4\chi_{ij} &= \phi_{iijj} - \sum_k
    \frac{\phi_{iik}\phi_{jjk}}{\omega_k^2} + \sum_k
    \frac{2\omega_k\left(\omega_i^2 + \omega_j^2 -
      \omega_k^2\right)\phi_{ijk}^2}{\Delta_{ijk}}\\ \Delta_{ijk} &=
    \left(\omega_i + \omega_j - \omega_k\right)\left(\omega_i +
    \omega_j + \omega_k\right)\left(\omega_i - \omega_j +
    \omega_k\right)\left(\omega_i - \omega_j - \omega_k\right)
\end{align}
From the resulting anharmonic constants $\chi$ the anharmonic
fundamental frequencies $\nu_i$ are obtained according to
\begin{align}
  \nu_i = \omega_i + 2\chi_{ii} + \frac{1}{2}\sum_{i\neq j} \chi_{ij}.
\label{eq:vpt2_freq}
\end{align}
In addition, overtones, combination bands and zero-point energies can
be determined, see e.g. Ref.~\citenum{yu2015vibrational}.\\

\noindent
In this work, the generalized VPT2\cite{barone2005anharmonic}
implementation (GVPT2) in the Gaussian\cite{g09} suite is used to
determine anharmonic frequencies. Specific settings including which
VPT2 model to use, step-size for numerical differentiation and
resonance thresholds are retained at their default values. Gaussian's
standardized interface initialized by the ``External'' keyword is used
to call an external script (i.e. the PhysNet potential), which
produces an energy, gradients and Hessians for a given
molecule/geometry. The output from the external potential is recovered
from a standard text file by Gaussian where the the cubic and quartic
terms are determined by numerical derivatives and used for a VPT2
calculation.\\

\section{Results}
\label{sec:Results}
The performance of the ML models is assessed by three points:
i) Out-of-sample errors for energies ($\Delta E$) and forces ($\Delta F$)
for which geometries of a separate test set are used. These errors
quantify to what extent the ML models are capable of interpolating
between training points. 
ii) Comparing harmonic frequencies determined
from the trained PhysNet model and conventional \textit{ab initio}
harmonic frequencies calculations at the same level of theory. These
harmonic frequencies also affect the anharmonic (VPT2) frequencies,
see Eq.~\ref{eq:vpt2_freq}), and
iii) Comparing VPT2 frequencies from the
PhysNet model with those from Gaussian at the MP2 level of theory and
with those from experiment, respectively. The comparison with
\textit{ab initio} MP2 VPT2 frequencies quantifies how closely the
PhysNet PES reproduces the \textit{ab initio} potential around the
minimum.\\

\subsection{Out-of-sample errors}
\begin{figure}[h!]
\centering \includegraphics[width=0.9\textwidth]{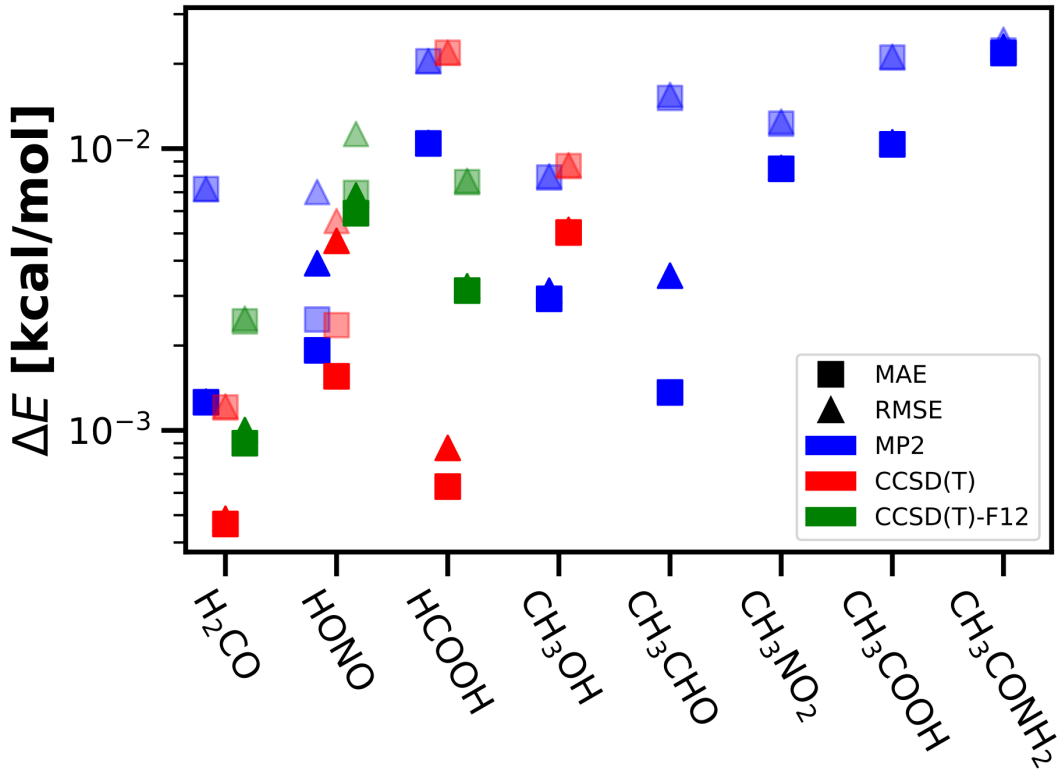}
\caption{The out-of-sample MAEs (squares) and RMSEs (triangles) of the
  energy for the different molecules. The levels of theory are
  color-coded and opaque and transparent symbols represent two PhysNet
  models trained independently on the same data. A general trend
  showing an increased error with increasing system size is
  visible. The lowest MAE is $\sim 0.0005$~kcal/mol (H$_2$CO, CCSD(T))
  and the highest is $\sim 0.0218$~kcal/mol (CH$_3$CONH$_2$, MP2). All
  PhysNet models predict the independent test set with chemical
  accuracy (better than 1~kcal/mol) and all out-of-sample performance
  measures are summarized in Table~S2.}
\label{fig:energy_learning}
\end{figure}

\begin{figure}[h!]
\centering \includegraphics[width=0.9\textwidth]{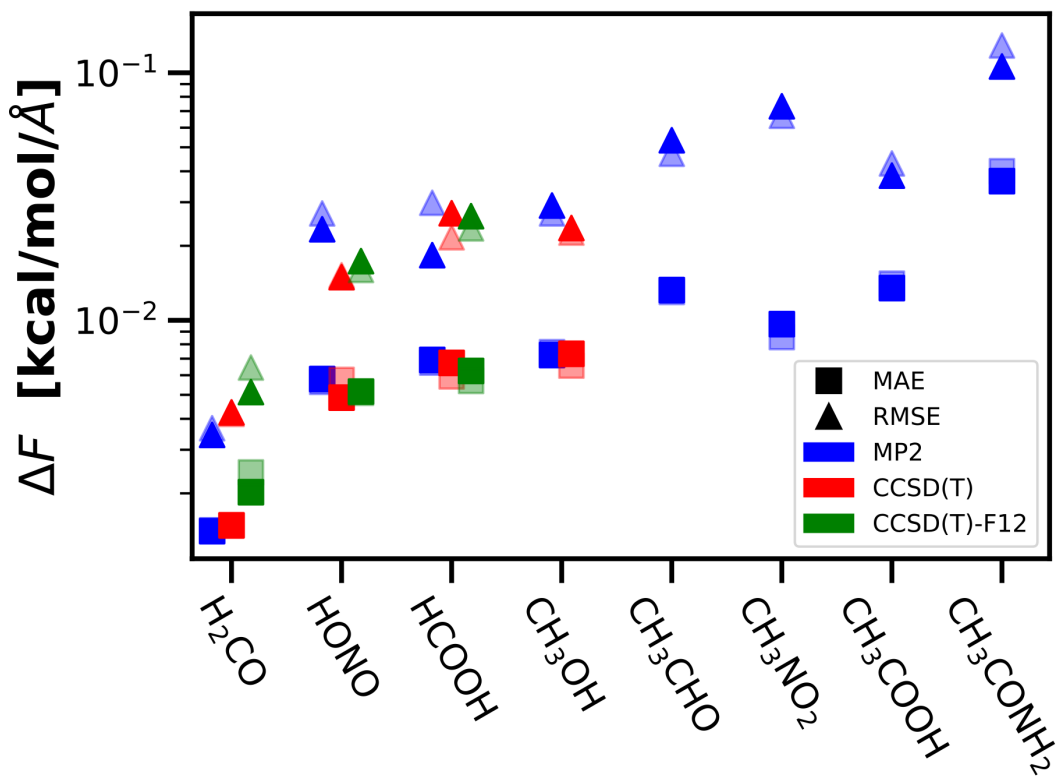}
\caption{The out-of-sample MAEs (squares) and RMSEs (triangles) of the
  forces for the different molecules. The levels of theory are
  color-coded, and the opaque and transparent symbols represent two
  PhysNet models trained independently on the same data. A general
  trend showing an increased error with increasing system size is
  visible. The lowest MAE is $\sim 0.0013$~kcal/mol/\r{A} (H$_2$CO,
  MP2) and the highest is $\sim 0.0400$~kcal/mol/\r{A}
  (CH$_3$CONH$_2$, MP2). All out-of-sample performance measures can be
  found in Tab.~S2.}
\label{fig:force_learning}
\end{figure}

All the PhysNet models were evaluated on separate test sets by means of
MAEs and root mean squared errors (RMSEs) for energies and forces
between reference calculations and predictions by PhysNet, see
Figures~\ref{fig:energy_learning} and \ref{fig:force_learning}. MAEs
(squares) and the RMSEs (triangles) at different levels of theory are
color coded. For each of the molecules and levels of theory two
independent PhysNet models (opaque and transparent symbols) were
trained on the same data to assess consistency and
reproducibility. The lowest MAE($E$) is $\approx 0.0005$~kcal/mol (for
H$_2$CO) whereas the largest is $\approx 0.0218$~kcal/mol (for
CH$_3$CONH$_2$) indicating that errors scale with system size. These
MAEs correspond to about 0.003 and 0.03~\% of the energy range spanned
by the data sets (16.3~kcal/mol for H$_2$CO and 72.6~kcal/mol for
CH$_3$CONH$_2$), respectively, and all $R^2$ coefficients are above
0.99998 (see Table~S2). From
Figure~\ref{fig:energy_learning} it is apparent that two independently
trained models (opaque and transparent symbols on the same vertical
line) can still differ appreciably (MAEs($E$) of $\sim 0.0013$ and
$\sim 0.0072$~kcal/mol for H$_2$CO trained on MP2 data, see
Tab.~S2) although both models are still of remarkable
quality.\\

\noindent
The appreciable variations between the two models trained on the same
data (opaque and transparent symbols, respectively) are notably
smaller for the force errors (i.e.  the opaque and transparent squares
are very close to each other), MAE($F$) and RMSE($F$), see
Figure~\ref{fig:force_learning}. Earlier work\cite{meuwly2020ml}
showed that this is a consequence of the different weights
(hyperparameters) in the loss function of PhysNet. During training,
the higher weight of the force leads to a slight deterioration of the
energies while the forces are still improving. The higher weight on
the forces, however, is motivated by the fact that accurate
derivatives are required for the present work. Similar to the errors
in the energies $\Delta E$, the force errors $\Delta F$ tend to
increase with increasing system size. The force errors for H$_2$CO are
notably lower compared to the other molecules. Because PhysNet is
invariant with respect to permutation of equivalent atoms, the C$_{\rm
  2v}$ symmetry of H$_2$CO could be responsible for the lower
out-of-sample errors. All test sets are predicted with energy errors
considerably better than chemical accuracy and, according to earlier
studies using PhysNet, are within the expected range (see
e.g. Refs.~\citenum{meuwly2020ml} and \citenum{mm.ht:2020} for H$_2$CO
and for larger molecules, respectively).\\

\subsection{Harmonic frequencies}
\begin{figure}[h!]
\centering \includegraphics[width=1.0\textwidth]{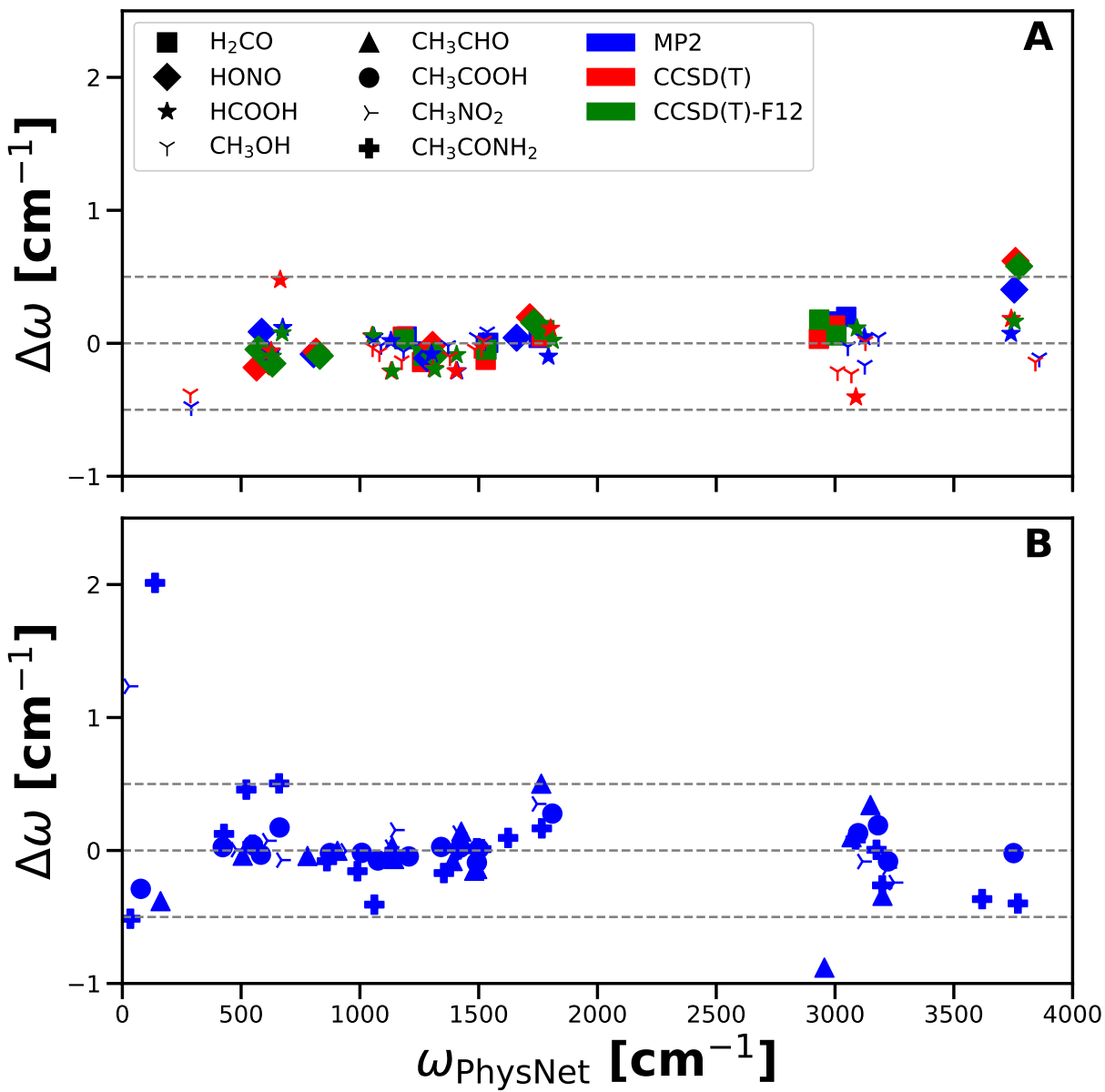}
\caption{The accuracy of the PhysNet harmonic frequencies is shown
  with respect to the appropriate reference \textit{ab initio}
  values. Here, $\Delta\omega$ corresponds to $\omega_{\rm
    ref}-\omega_{\rm PhysNet}$ and the figure is divided into two
  windows for clarity. For the small molecules (A), irrespective of
  the level of theory, all frequencies are reproduced to within less
  than 1~cm$^{-1}$. For the larger molecules (B), deviations larger
  than 1~cm$^{-1}$ are found. Largest errors are found for
  CH$_3$NO$_2$ (1.23~cm$^{-1}$) and for CH$_3$CONH$_2$
  (2.01~cm$^{-1}$). Tables containing the PhysNet and \textit{ab
    initio} harmonic frequencies can be found in
  Tabs.~S3-S5,
  S7-S9,
  S11-S13,
  S15, S16,
  S18, S20,
  S22, S24.}
\label{fig:harm_freq}
\end{figure}

Harmonic frequencies computed from the PhysNet representations are
useful to test how well the ML model reproduces the region close to
the minimum of the PES compared with frequencies determined directly
from the \textit{ab initio} calculations. Also, they provide the
anharmonic frequencies through Eq. \ref{eq:vpt2_freq} which will be
essential for the calculation of VPT2
frequencies. Figure~\ref{fig:harm_freq} compares the reference
\textit{ab initio} harmonic frequencies from MOLPRO\cite{MOLPRO} with
those from PhysNet after optimizing the structure of each molecule
with the respective energy function.  From the two models trained
independently on the same data, only the frequencies from the best
PhysNet model are discussed. The complete list of PhysNet and
\textit{ab initio} harmonic frequencies for all molecules and at all
levels of theory are reported in
Tables~S3-S5,
S7-S9,
S11-S13,
S15, S16,
S18, S20,
S22, and S24.\\

\noindent
For the smaller molecules (Figure~\ref{fig:harm_freq}A), all harmonic
frequencies from the PhysNet models are within 1~cm$^{-1}$ (mostly $<
0.5$~cm$^{-1}$) of the reference {\it ab initio} calculations and the
MAEs($\omega$) range from 0.08 to 0.21~cm$^{-1}$. Similarly, for the
larger molecules (Figure~\ref{fig:harm_freq}B) most of the harmonic
frequencies are within 1~cm$^{-1}$ of the reference values. Single
larger differences are found for the low frequency modes with errors
of 1.23 and 2.01~cm$^{-1}$ for CH$_3$NO$_2$ and CH$_3$CONH$_2$,
respectively. The remaining PhysNet frequencies reproduce the {\it ab
  initio} harmonic frequencies with errors smaller than 1~cm$^{-1}$
and MAEs($\omega$) are between 0.09 and 0.28~cm$^{-1}$. Note that all
the optimized structures are true minima (i.e. no imaginary
frequencies were found).\\

\subsection{VPT2 frequencies}
\begin{figure}[h!]
\centering
\includegraphics[width=1.0\textwidth]{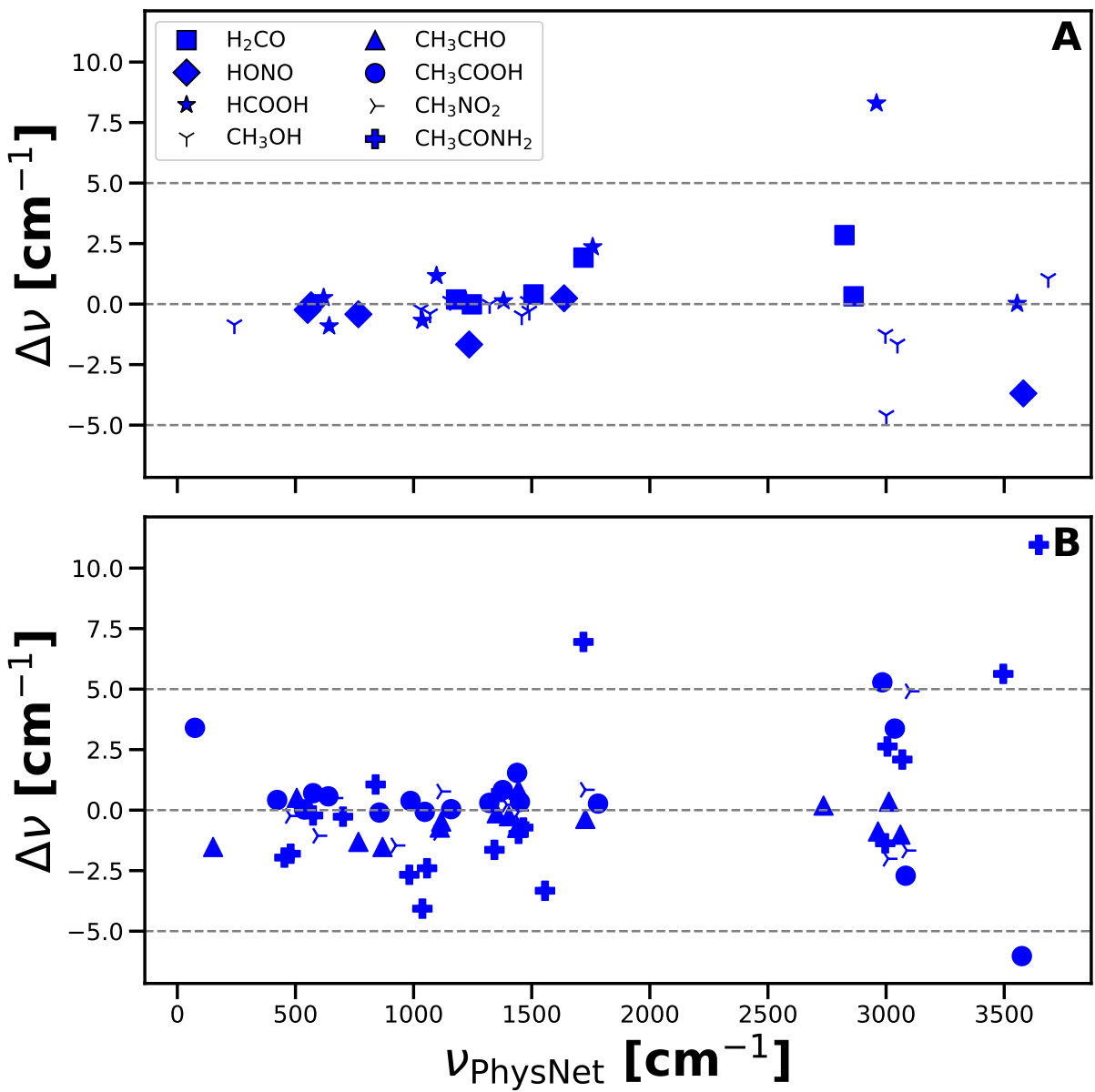}
\caption{The accuracy of the PhysNet MP2 VPT2 frequencies is shown
  with respect to their MP2 \textit{ab initio} values. Here,
  $\Delta\nu$ corresponds to $\nu_{\rm MP2}-\nu_{\rm PhysNet}$ and the
  figure is divided into two windows for clarity. For most of the
  molecules and frequencies, the reference anharmonic frequencies are
  reproduced with deviations smaller than $10$~cm$^{-1}$ (or even
  5~cm$^{-1}$ for the small molecules). The anharmonic frequencies of
  CH$_3$NO$_2$ and CH$_3$CONH$_2$ having negative frequencies are not
  shown. Tables containing the PhysNet and \textit{ab initio} VPT2
  frequencies can be found in
  Tabs.~S6,
  S10,
  S14,
  S17,
  S19, S21,
  S23, S25.}
\label{fig:mp2_anharm_freq}
\end{figure}

Next the quality of the VPT2 frequencies for the best (based on the
MAE($\nu$)) PhysNet models is assessed. For this, VPT2 frequencies
computed from PhysNet models trained on MP2 data are compared to their
\textit{ab initio} counterparts, see
Figure~\ref{fig:mp2_anharm_freq}. This comparison was not done for the
higher levels of theory because the VPT2 method in Gaussian is only
available for levels with analytical second derivatives\cite{g09} and
because the calculations get very expensive for the larger
molecules. The majority of the MP2 VPT2 frequencies are reproduced to
within better than 5~cm$^{-1}$ with single larger differences of up to
$\sim 10$~cm$^{-1}$. It is apparent that the PhysNet VPT2 frequencies
for the smaller molecules (Figure~\ref{fig:mp2_anharm_freq}A) are more
accurate - with MAEs($\nu$) between 0.95 and 1.55~cm$^{-1}$ - than the
frequencies of the larger molecules
(Figure~\ref{fig:mp2_anharm_freq}B). For the larger molecules
MAEs($\nu$) between 1.14 and 2.71~cm$^{-1}$ are found. The good
agreement between \textit{ab initio} and PhysNet VPT2 frequencies
further confirms the high quality of the ML potentials as third and
fourth order derivatives of the potential are very sensitive to the
shape and accuracy of the PES. Note that for two molecules
(CH$_3$NO$_2$ and CH$_3$CONH$_2$) one and two negative VPT2
frequencies are found for the lowest frequency modes from PhysNet as
well as from the \textit{ab initio} calculations. This is a deficiency
of VPT2 and can be explained by Equation~\ref{eq:vpt2_freq}. If the
perturbation $2\chi_{ii} +\frac{1}{2}\sum_{i\neq j} \chi_{ij}$ exceeds
the harmonic frequency $\omega_i$ (and is negative), which can occur
in particular for small $\omega_i$, the VPT2 frequency can become
negative. These frequencies are reported in
Tables~S23 and S25. Even
though a direct comparison of the higher level PhysNet VPT2 to the
\textit{ab initio} reference is intractable, it seems reasonable to
assume that the coupled cluster quality PhysNet models reach similar
accuracies. This is further supported by the findings for the harmonic
frequencies at the CCSD(T) and CCSD(T)-F12 levels.\\

\begin{figure}[h!]
\centering \includegraphics[width=1.0\textwidth]{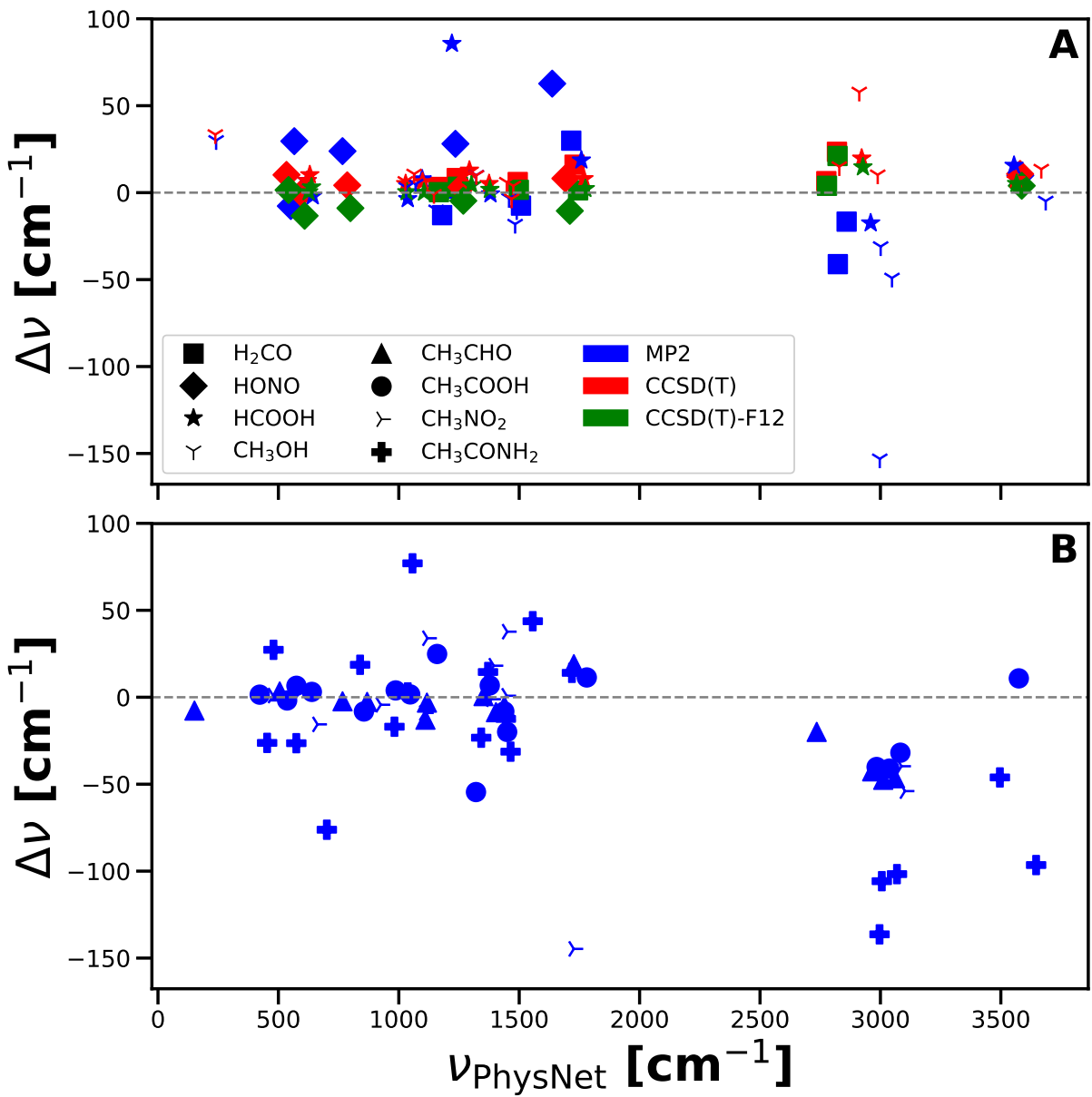}
\caption{VPT2 frequencies from PhysNet compared with experimental
  values\cite{herndon2005determination, guilmot1993rovibrational,
    tew2016ab, serrallach1974methanol, wiberg1995acetaldehyde,
    goubet2015standard, wells1941infra, ganeshsrinivas1996simulation}.
  Here, $\Delta\nu = \nu_{\rm exp}-\nu_{\rm PhysNet}$ and the figure
  is divided into two panels for smaller (panel A) and larger (panel
  B) molecules. For H$_2$CO good agreement is achieved with
  experiment, where the frequencies converge towards the experiment
  going from MP2 to CCSD(T)-F12 with a maximum deviation of $\sim
  20$~cm$^{-1}$. A similar trend is also visible for HCOOH and
  CH$_3$OH. In panel B, larger deviations are visible especially for
  high frequencies. Tables
  S6,
  S10,
  S14,
  S17,
  S19, S21,
  S23, S25 report the
  PhysNet, MP2 \textit{ab initio} VPT2, and experimental frequencies.
  For HCOOH the literature suggests different frequencies for the OH
  bending mode, see Refs.~\cite{millikan1957fam,tew2016ab} (1229 and
  1302~cm$^{-1}$, respectively). Here, a frequency of 1302 is used,
  following Ref.~\citenum{tew2016ab}.}
\label{fig:anharm_freq}
\end{figure}

\noindent
The (anharmonic) PhysNet VPT2 frequencies can also be compared
directly to experimental data, see Figure~\ref{fig:anharm_freq}. For
the small molecules good agreement is found and the frequencies
converge towards the experimental values when going from MP2 to
CCSD(T)-F12. VPT2 frequencies from the PhysNet model for H$_2$CO
trained on CCSD(T)-F12 data reproduce the experimental
data\cite{herndon2005determination} with a maximum deviation of $\sim
20$~cm$^{-1}$ and a MAE($\nu$) of $\sim 4$~cm$^{-1}$. Similar trends
are also visible for the other CCSD(T)-F12 models where a MAE($\nu$)
of $\sim 7$~cm$^{-1}$ ($\sim 4$~cm$^{-1}$) is found for HONO (HCOOH),
see Table~\ref{tab:all_mae_vpt2_exp}. In going from MP2 to CCSD(T)-F12
the MAE reduces by a factor of almost five. For CH$_3$OH the highest
quality PhysNet model is trained on CCSD(T) reference
calculations. Here, larger deviations of up to $\sim 58$~cm$^{-1}$ for
single vibrations and a MAE($\nu$) = 14.30~cm$^{-1}$ are found.\\

\noindent
For the larger molecules (Figure~\ref{fig:anharm_freq}B) the PhysNet
models are only trained on MP2 reference data. As judged from the
improvements between VPT2 calculations at the MP2, CCSD(T), and
CCSD(T)-F12 levels with experiment for H$_2$CO, HONO and HCOOH, for
PhysNet trained on MP2 data only modest agreement between computed and
experimentally observed frequencies is expected for the larger
molecules. VPT2 modes from the PhysNet models trained on MP2 data for
the larger molecules (Figure~\ref{fig:anharm_freq}B) with frequencies
below $\sim 2000$~cm$^{-1}$ are centered around the experimental
values with a MAE($\nu$) between 6 and 30~cm$^{-1}$. The higher
frequencies, however, tend to be overestimated by PhysNet trained on
data at the MP2 level of theory (MAE($\nu$) between 31 and
97~cm$^{-1}$). Because VPT2 frequencies from PhysNet and \textit{ab
  initio} MP2 calculations agree well (see
Figure~\ref{fig:mp2_anharm_freq}) it is concluded that the (large)
errors are not caused by the ML method itself but rather are a
deficiency of the MP2 level of theory and/or of the VPT2 approach when
compared with experiment. Nevertheless, for completeness, the MAEs
with respect to experiment are given in
Table~\ref{tab:all_mae_vpt2_exp}. As it becomes computationally
unfeasible to calculate a comprehensive CCSD(T) data set containing
some thousand data points for the larger molecules, an alternative
approach to improve the ML models is required. This is explored
next.\\

\subsection{Transfer learning to coupled cluster quality}
As the accuracy of the MP2 based models is limited and as it is
computationally too expensive to obtain comprehensive coupled cluster
quality training data sets for the larger molecules, alternative
methods have to be considered. Among NN based methods TL can be used
to exploit correlations between data at different levels of theory
\cite{smith2018outsmarting}. TL uses the knowledge, acquired when
learning how to solve a task A, as a starting point to learn how to
solve a related, but different task
B\cite{pan2009survey,taylor2009transfer}. Methods related to TL are
$\Delta$-machine learning~\cite{DeltaPaper2015} or multi-fidelity
learning~\cite{batra2019multifidelity} which become increasingly
popular in quantum chemical applications
\cite{smith2018outsmarting,mm.ht:2020,nandi2021delta,mezei2020noncovalent}. Also,
TL bears similarities with potential morphing techniques which exploit
the fact that the overall shapes of PESs at sufficiently high levels
of theory are related and can be transformed by virtue of suitable
coordinate
transformations\cite{meuwly1999morphing,bowman1991simple}.\\

\noindent
Here, TL is used to obtain coupled cluster quality PESs for the larger
molecules based on models trained at the MP2 level. For this, the MP2
models are used as a good initial guess and only a fraction of the
data set, calculated at the CCSD(T) level, needs to be provided for
TL. To transfer learn the models, the parameters of a PhysNet model
trained on MP2 data is used to initialize the PhysNet training. Then,
PhysNet is trained (i.e. the parameters are adjusted) using energies,
forces and dipole moments from the higher level of theory. The
training is performed following the same approach and using the same
hyperparameters as when training a model from scratch, except that the
learning rate is decreased to values between $10^{-5} - 10^{-4}$.\\

\begin{figure}[h]
\centering \includegraphics[width=1.0\textwidth]{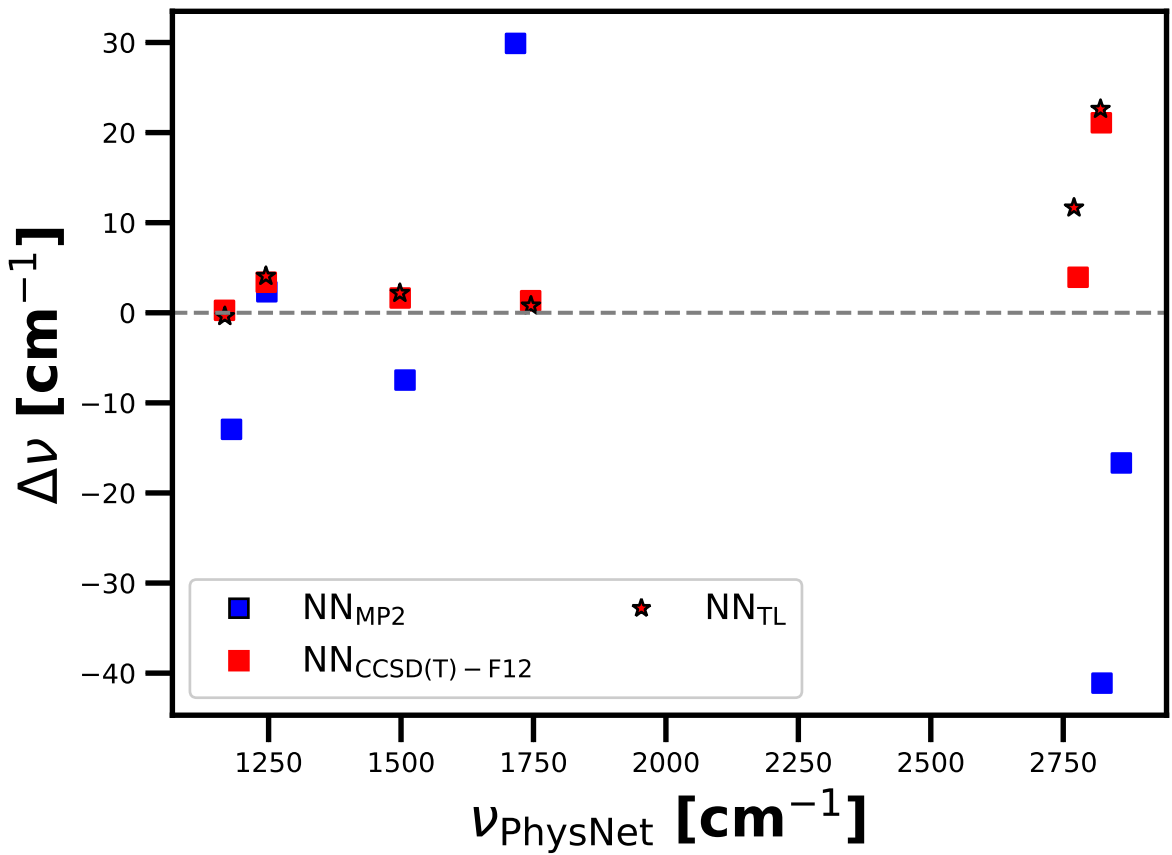}
\caption{Comparison of VPT2 frequencies from PhysNet trained on MP2
  (NN$_{\rm MP2}$), CCSD(T)-F12 (NN$_{\rm CCSD(T)-F12}$) and TL
  (NN$_{\rm TL}$) from MP2 to CCSD(T)-F12 with the experiment
  \cite{herndon2005determination} for H$_2$CO. Here, $\Delta\nu =
  \nu_{\rm exp}-\nu_{\rm PhysNet}$. The VPT2 frequencies on the TL-PES
  are within 8~cm$^{-1}$ to the values obtained from a model trained
  from scratch yielding an improvement in comparison to an MP2 model
  (its ``starting point'').}
\label{fig:tl_h2co}
\end{figure}

\noindent
First, TL is applied to H$_2$CO which serves as a ``toy model''
because PhysNet models at different levels of theory are readily
available and allow for direct comparison. In other words, the TL
model based on MP2 data and retrained with input from CCSD(T)-F12 can
be directly compared with PhysNet trained entirely on CCSD(T)-F12
reference data. For this, the MP2 model is used as initial guess and
transfer learned to CCSD(T)-F12 quality using 188 data points (151 of
the original data set extended with 37 VPT2 geometries) and tested on
the remaining 3450 structures from the original CCSD(T)-F12 data
set. The TL model shows slightly lower errors for energies (MAE($E$) =
0.0004~kcal/mol) compared to the model trained from scratch (MAE($E$)
= 0.0009~kcal/mol)), whereas the forces are slightly less accurate
with MAE($F$) of 0.0066 and 0.0020~kcal/mol/{\AA}  for TL and from
scratch, respectively. For the harmonic frequencies, all reference
CCSD(T)-F12 values are reproduced with deviations smaller than
0.3~cm$^{-1}$ and with a MAE($\omega$) = 0.1~cm$^{-1}$, which is
similar for a model trained from scratch, see
Tables~S5 and
S26. The most relevant measure of
performance is the direct comparison of VPT2 to experimental
frequencies, especially when the high level \textit{ab initio}
calculation of harmonic frequencies become too
expensive. Figure~\ref{fig:tl_h2co} shows the deviation of a transfer
learned PhysNet model, a PhysNet model trained on CCSD(T)-F12 data and
a model trained on MP2 data (the ``base model'' for TL) with respect
to the experimental H$_2$CO frequencies. VPT2 frequencies from
NN$_{\rm TL}$ are very close to those from NN$_{\rm CCSD(T)-F12}$
except for one high frequency mode and clearly superior to those from
NN$_{\rm MP2}$ throughout when compared with experiment.\\

\noindent
TL is used to obtain high quality PESs for all the models for
which only MP2 reference data are available (i.e. CH$_3$CHO,
CH$_3$NO$_2$, CH$_3$COOH and CH$_3$CONH$_2$). For each of the
molecules around 5~\% or less of the original, lower level, data set
is recalculated at the CCSD(T) level. The TL data sets contain
geometries sampled using the normal mode approach (evenly split among
the different temperatures), VPT2 geometries and the MP2 optimized
geometry. They contained CCSD(T) information (energies, gradients and
dipole moments) on 262, 452, 542 and 632 CCSD(T) geometries for
CH$_3$CHO, CH$_3$NO$_2$, CH$_3$COOH and CH$_3$CONH$_2$,
respectively. The TL data sets are again split randomly according to
85/10/5~\% into training/validation/test sets for TL. As a
consequence, the TL models are tested only on a small number of
molecular geometries. \\

\noindent
For the larger molecules assessing the performance of a TL model is
more difficult. On the one hand, the rather small number of geometries
in the TL set are mostly used for training and validation. Thus, only
a small fraction ($< 50$ geometries) remains for testing the models
and might give less meaningful results statistically. The evaluation
of the separate test set for all molecules yields MAEs($E$) smaller
than 0.03~kcal/mol and MAE($F$) smaller than 0.02~kcal/mol/{\AA} (compare
to Figure~\ref{fig:energy_learning} and \ref{fig:force_learning}) and
are within the realm of what was expected. The MAEs and RMSEs of
energies and forces for all TL models are listed in
Table~S28 for completeness.\\

\noindent
A second measure of performance are the harmonic frequencies which can
be compared with those from direct \textit{ab initio} calculations.
Even though optimizations and normal mode calculations at the CCSD(T)
level become computationally expensive rather quickly for larger
molecules, they were performed for all the larger molecules CH$_3$CHO,
CH$_3$NO$_2$, CH$_3$COOH and CH$_3$CONH$_2$. For CH$_3$CHO,
CH$_3$NO$_2$ and CH$_3$COOH the harmonic frequencies on the TL PhysNet
models compare to within MAE($\omega$) = 0.2~cm$^{-1}$, MAE($\omega$)
= 0.3~cm$^{-1}$ and MAE($\omega$) = 0.1~cm$^{-1}$ with the explicit
calculations at the CCSD(T) level of theory, see
Tables~S29 to
S31. These errors are also within those
achieved by the models trained from scratch. Slightly larger errors
are found for CH$_3$CONH$_2$ (MAE($\omega$) = 1.1~cm$^{-1}$, see
Table~S32).\\

\begin{figure}[h]
\centering \includegraphics[width=1.0\textwidth]{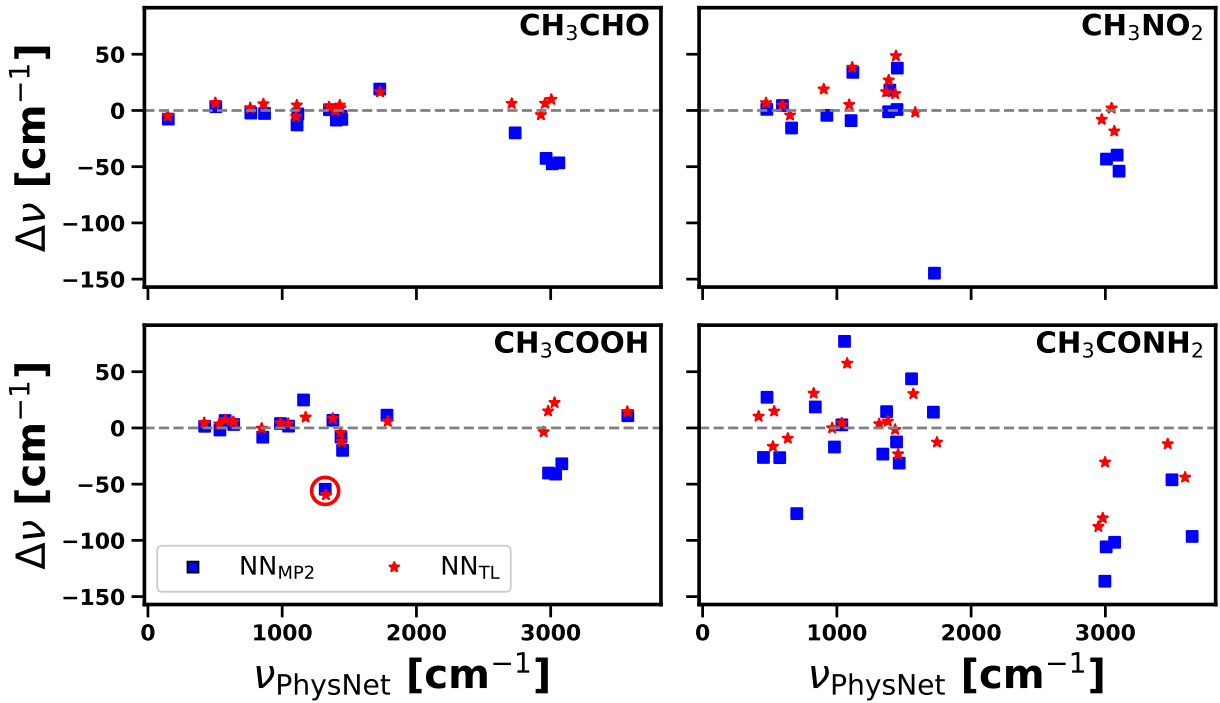}
\caption{Comparison of the VPT2 frequencies from PhysNet trained on
  MP2 (NN$_{\rm MP2}$, blue squares) and TL to CCSD(T) (NN$_{\rm TL}$,
  red asterisks) with experimental values. Here, $\Delta\nu = \nu_{\rm
    exp}-\nu_{\rm PhysNet}$. The TL models yield overall improved VPT2
  frequencies, especially for the high frequency modes. A rather large
  deviation is found for the frequency around 1300~cm$^{-1}$ for
  NN$_{\rm TL}$ for CH$_3$COOH. A red circle marks a surprisingly high
  error, compared to the other modes. It is possible that this
  discrepancy is caused by a misassignment of experimental modes, see
  text.}
\label{fig:tl_large_molecules}
\end{figure}

\noindent
Finally, the performance of the models transfer learned to CCSD(T) can
be assessed by comparing the VPT2 frequencies with experiment, see
Figure~\ref{fig:tl_large_molecules}. As a reference, the VPT2
frequencies from PhysNet models trained on MP2 data (blue squares) are
reported together with the TL models. It is apparent that the TL
models are closer to experiment especially for the high frequency
modes and all MAEs($\nu$) with respect to experiment are reduced, see
Table~\ref{tab:all_mae_vpt2_exp}. For CH$_3$CHO
the MAE($\nu$) is reduced from 15.3 to 5.5 cm$^{-1}$, corresponding to
a factor of almost 3, while for the remaining three molecules this
factor is closer to 2. Good agreement is found for CH$_3$CHO,
CH$_3$NO$_2$ and CH$_3$COOH while the picture is less clear for
CH$_3$CONH$_2$.\\

\begin{table}[h]
\begin{tabular}{lccc|c}
\toprule
 \textbf{MAE($\nu$) [cm$^{-1}$]} & \textbf{MP2}&  \textbf{CCSD(T)} & \textbf{CCSD(T)-F12} & \textbf{scaled MP2/CCSD(T)}\\
 \midrule
H$_2$CO & 18.40 & 9.04 & 3.98/{\bf 6.95} & 48.22/37.74\\
HONO & 26.98 & 5.56 & 6.74 & 38.44/17.31\\
HCOOH & 17.61 & 9.06 & 3.59 & 32.99/26.31\\
CH$_3$OH & 27.28 & 14.30 & -- & 23.47/19.60 \\
CH$_3$CHO	&	15.27	&	{\bf 5.47} & -- & 29.04/21.90\\
CH$_3$COOH	&	16.25	&	{\bf 10.83} &-- & 21.39/19.74	\\
CH$_3$NO$_2$	&	28.56	&{\bf 15.38} &	-- & 34.27/25.28 \\
CH$_3$CONH$_2$	&	47.23	&{\bf 25.13} &-- & 43.46/34.81\\
\bottomrule
\end{tabular}
\caption{MAEs of the PhysNet VPT2 frequencies with respect to
  experimental values\cite{herndon2005determination,
    guilmot1993rovibrational, tew2016ab, serrallach1974methanol,
    wiberg1995acetaldehyde, goubet2015standard, wells1941infra,
    ganeshsrinivas1996simulation}. The level of theory of the training
  data is given in the header and the results of the transfer learned
  PhysNet models are highlighted in bold. An additional column shows
  the MAEs of the ``conventional'' approach where harmonic frequencies
  are scaled by an empirical factor\cite{cccbdb} (see also
  Figure~S1 for an illustration of the scaled
  frequencies compared with experiment).}
\label{tab:all_mae_vpt2_exp}
\end{table}

\subsection{Timings}
By using \textit{ab initio} calculations (see Table~S1 for cost of \textit{ab initio} energy, forces and dipole
moment calculations) and a PhysNet representation of the underlying
PES, it is possible to determine VPT2 frequencies at considerably
higher levels of theory for larger molecules than is possible from
straight {\it ab initio} evaluations. This raises the question: how do
the computational efficiencies of the different approaches compare?\\

\noindent
The following points are considered:

\begin{itemize}
\item {\it Feasibility at a given \textit{ab initio} level:} VPT2
  calculations at the MP2 level have been performed for all molecules
  considered. The calculation times ranged from few CPU hours for the
  small molecules (H$_2$CO: 2.5~h, HONO: 5~h), to multiple CPU hours
  for HCOOH/CH$_3$OH (14~h) and CH$_3$CHO (60~h) up to several days
  for the largest molecules (CH$_3$COOH: 190~h, CH$_3$NO$_2$: 103~h,
  CH$_3$CONH$_2$:~298~h). However, compared with experiment, VPT2
  frequencies at the MP2 level of theory were found to be inferior to
  CCSD(T) with unsigned deviations of up to $~150$~cm$^{-1}$, see
  Table~\ref{tab:all_mae_vpt2_exp}. While a full CCSD(T) VPT2
  treatment for the smallest molecules would be possible, this will be
  too costly for the larger structures. It was
  reported\cite{jacobsen2013anharmonic} that a VPT2 calculation for
  H$_2$CO at various levels of theory is around 10 times more
  expensive than the calculation of harmonic frequencies. This is also
  found in the present work. Even though this factor will increase for
  larger molecules (as more derivatives need to be evaluated) it can
  be used to tentatively assess the cost of a CCSD(T) VPT2 treatment
  of a larger molecule. The calculation of harmonic frequencies of
  CH$_3$CONH$_2$ at the CCSD(T) level is found to take more than 20
  CPU days which suggests that a CCSD(T) VPT2 calculation would take
  more than 200 CPU days.\\

\item {\it Data Generation and Learning of PhysNet:} The time for
  generating a PhysNet model includes computation of a data set
  (\textit{ab initio} calculations of energies, forces and dipole
  moments), training and validating the NN. The most costly steps are
  the training ($\leq 5$~days) and, depending on the level of theory,
  the calculation of the reference data. As an example, generating the
  most expensive data set (5401 HCOOH geometries at the CCSD(T)-F12
  level: $\sim 130$~CPU min per calculation) takes around two days
  assuming that 250 calculations can be performed in parallel (the
  ``real time'' of the calculations might be longer depending on I/O
  performance). The PhysNet VPT2 calculation then is the least
  expensive step, taking less than 1~hour (H$_2$CO: 18 min,
  CH$_3$CONH$_2$: 58 min).\\

  \noindent
  As an explicit example, data generation at the MP2 level, NN
  training and VPT2 calculations for CH$_3$COOH or CH$_3$CONH$_2$ are
  considered. With a conservative estimate  for generating
    reference data, fitting and testing of the PhysNet model of 7
    days, a direct \textit{ab initio} VPT2 calculation at the MP2
  level of theory exceeds the cost of the NN + VPT2 approach for
  these two molecules by a factor of 1.1 (10 \%) and 1.7 (70 \%),
  respectively. For larger molecules this factor increases further as
  it does also for higher-level methods. In addition, the approach
  pursued here has the added benefit of a full-dimensional,
  near-equilibrium PES that can also be used for
  MD simulations, albeit some additional reference calculations may be
  required.\\

\item {\it Crossover between NN + VPT2 versus ab initio:} It is of
  interest to ask at what point does the NN + VPT2 approach become
  advantageous over the \textit{ab initio} approach in terms of
  overall computational cost. For the higher levels of theory the cost
  of \textit{ab initio} VPT2 calculations scales less favourably than
  single (energy, forces and dipole moment) calculations due to the
  repetitive evaluation of first- to fourth-order derivatives of the
  potential energy. If the cost for an \textit{ab initio} VPT2
  calculation is estimated\cite{jacobsen2013anharmonic} as $10*t_{\rm
    harm}$ (where $t_{\rm harm}$ is the time to calculate harmonic
  frequencies), then a PhysNet model becomes more favourable already
  for HCOOH at the CCSD(T) level ($t_{\rm harm} = 30$~h). The
  preference of the NN + VPT2 approach for larger molecules and higher
  levels is even more apparent when using TL. As an example, the
  CCSD(T) data set for CH$_3$CONH$_2$ (632 data points; the
  calculations including energy, forces and dipole moments take $\sim
  9$ CPU hours on average while the real time was closer to 15~h) is
  generated in less than 2 days, and the TL can be performed with
  costs on the order of 1~hour. Together with the time for reference
  data generation at the MP2 level, training and evaluating of the
  base NN, the total estimated time for VPT2 frequencies for
  CH$_3$CONH$_2$ at the CCSD(T) level on the order of 10 CPU days,
  compared with an estimated 200 CPU days from a brute force
  calculation.  Empirically, this is close to linear scaling, whereas
  at the \textit{ab initio} level the scaling increases formally from
  $N^5$ for MP2 to $N^7$ for CCSD(T)\cite{friesner2005ab}.\\
\end{itemize}

\noindent
In summary, while at the MP2 level the NN + VPT2 approach is
computationally superior only for the two largest molecules, it
outperforms the \textit{ab initio} approach at higher levels for all
but tetra-atomic molecules. The feasibility of the NN + VPT2
approach for larger molecules and higher levels is strengthened
when employing TL.\\

\section{Discussion}
The results presented so far have established that a ML model
yields the same normal mode and VPT2 frequencies as direct evaluations
from electronic structure calculations and comparison between
computations and experiments is favourable (unsigned deviations
between $< 0.5$ and 21~cm$^{-1}$ for the CCSD(T)-F12 models and
between $< 0.5$ and 88~cm$^{-1}$ for the CCSD(T) models, including
TL). Alternatively, calculated normal mode frequencies are often
scaled by empirical factors to compare directly with
experiment.\cite{scott1996harmonic} Table~\ref{tab:all_mae_vpt2_exp}
summarizes the mean absolute errors between experimentally measured
frequencies, those from VPT2 calculations at the three levels of
theory, and the scaled harmonic frequencies from MP2 and CCSD(T)
calculations (last column). Scaled MP2 and CCSD(T) frequencies differ
on average by 20 cm$^{-1}$ to 50 cm$^{-1}$ from those measured
experimentally. This is comparable to the difference between
(anharmonic) VPT2 calculations at the MP2 level on the PhysNet-PES and
experiment. On the other hand, VPT2 calculations on the CCSD(T) and
CCSD(T)-F12 PESs are in better agreement (average MAE of 10 cm$^{-1}$
and 5 cm$^{-1}$) and clearly outperform results from MP2 calculations
(scaled harmonic and VPT2) and scaled harmonic CCSD(T) frequencies.\\

\noindent
Application of TL showed that the harmonic frequencies of the TL
models reach MAE($\omega$) with respect to reference comparable to
models trained from scratch. The small additional cost for obtaining a
high-quality PES from a lower-level PES and the rapid scaling of the
costs for obtaining \textit{ab initio} harmonic frequencies motivates
the use of TL for the determination of high quality harmonic
frequencies for which fully-dimensional \textit{ab initio} PESs are
computationally too expensive.\\

\noindent
Comparison of accurate (anharmonic) VPT2 frequencies with experiment
can provide additional insight. Such comparisons depend on the quality
of the computations and that of the experiment. One particularly
appealing possibility is to compare and potentially reassign
individual modes. As an example, the OH bending vibration of monomeric
formic acid is considered. The early infrared studies were carried out
in the late 1950s by Milliken and Pitzer \cite{millikan1957fam} and by
Miyazawa and Pitzer \cite{miyazawa1959fam}, where the OH bending
vibration was assigned to a signal at 1229~cm$^{-1}$. Later work (see
e.g. Refs. \citenum{freytes2002overtone} or ~\citenum{reva1994ir} and
references therein) reported an OH bending frequency of 1223~cm$^{-1}$
consistent with the earlier work. An OH bending frequency of
1223~cm$^{-1}$ is $\sim 80$~cm$^{-1}$ lower than what is found from
VPT2 calculations on the PhysNet PES at the CCSD(T)-F12 level of
theory which finds it at 1302~cm$^{-1}$.\\

\noindent
Recent theoretical work determined vibrational
  frequencies from vibrational configuration interaction (VCI)
  calculations on a global CCSD(T)(F12*)/cc-pVTZ-F12 PES for cis- and
  trans-formic acid and tried to re-assign experimentally determined
  fundamental, overtone and combination
  bands.\cite{tew2016ab,freytes2002overtone} These calculations
proposed to assign the fundamental OH bend to an experimentally
measured frequency at 1306.2~cm$^{-1}$ and the first overtone of the
OH torsion to a frequency at 1223~cm$^{-1}$, converse to previous
assignments.\cite{freytes2002overtone} Due to the rather close
spacing, these two vibrations are in strong 1:2 Fermi resonance. This
assignment (see Figure~\ref{fig:anharm_freq}) is consistent with the
VPT2 calculations on the PhysNet(CCSD(T)-F12) PES results which find
the OH bend at 1301.9~cm$^{-1}$. In addition, the first overtone of
the OH torsion can also be calculated using PhysNet + VPT2 and
is found at 1220.2~cm$^{-1}$, only 3 cm$^{-1}$ away from an
experimentally determined signal. Given the agreement of the two
recent high-level computational treatments, based on very different
approaches, and the fact that experiment indeed finds vibrational
bands around 1220 and 1300 cm$^{-1}$, a reassignment of the
fundamental OH bend to a frequency of 1306.2 cm$^{-1}$ is supported.\\

\noindent
A similar situation arises for CH$_3$COOH
(Figure~\ref{fig:tl_large_molecules}, red circle), where the signal at
1325.5 ~cm$^{-1}$ from VPT2 calculation on the TL PhysNet(CCSD(T)) PES
disagrees by $\sim 60$~cm$^{-1}$ from the experimentally measured
frequency at 1266~cm$^{-1}$.\cite{goubet2015standard} While (mostly)
older
studies\cite{wilmshurst1956infrared,haurie1965spectres,berney1970infrared,maccoas2003rotational}
assigned the C-O stretch to a frequency around 1260~cm$^{-1}$ and the
OH bend to a lower frequency around 1180~cm$^{-1}$, more recent work
came to a converse assignment\cite{goubet2015standard,marechal1987ir}
($\nu_7= 1266$~cm$^{-1}$ and $\nu_8= 1184$~cm$^{-1}$, following the
notation in Reference~\citenum{goubet2015standard}). Goubet et
al.\cite{goubet2015standard} noted that $\nu_7$ and $\nu_8$ are not
well predicted by theoretical calculations at the anharmonic level
(RI-MP2/aVQZ harmonic frequencies corrected with VPT2 corrections
obtained at the B98/aVQZ level following $\nu_{\rm RI-MP2} =
\omega_{\rm RI-MP2} - \left(\omega_{B98} - \nu_{B98}\right)$) and
ascribed the discrepancy to Fermi resonances between $\nu_7$ and
$\nu_8$ with the first overtone of $\nu_{16}$ (A$''$ fundamental at
642~cm$^{-1}$) and $\nu_{11}$ (A$''$ fundamental at
581.5~cm$^{-1}$). Similarly, a large deviation of --60~cm$^{-1}$
between TL PhysNet(CCSD(T)) and $\nu_7$ is found in this study
compared with --61~cm$^{-1}$ in
Reference~\citenum{goubet2015standard}.\\

\noindent
Besides Fermi resonance as an explanation for the discrepancies,
little work has been done on overtone and combination
bands. Reference~\citenum{olbert2011raman}, however, suggested an
alternative assignment of two bands at 1324.4 and 1259.4~cm$^{-1}$
previously observed in the IR spectra of acetic acid isolated in Ar
matrix \cite{berney1970infrared,macoas2004photochemistry}. They
proposed to reassign the frequency at 1324.4~cm$^{-1}$ to $\nu_7$ and
1259.4~cm$^{-1}$ to the first overtone of $\nu_{16}$ based on VPT2
calculations at the B3LYP/6-311++G(2d,2p) level and deuteration
experiments. This new assignment is supported by the NN + VPT2 results
yielding frequencies of 1325.5 and 1250.2~cm$^{-1}$ for $\nu_7$ and
$2\nu_{16}$, respectively. These present findings, for formic acid and
acetic acid, encourage further theoretical work on overtones and
combination bands.\\

\section{Conclusion and Outlook}
\label{sec:Conclusion}
The combined NN + VPT2 approach together with TL is shown to provide
accurate anharmonic frequencies at high levels of electronic structure
theory. The PhysNet code was adapted to additionally predict
analytical derivatives of the dipole moment $\bm{\mu}$ and second
order derivatives of $E$ with respect to Cartesian coordinates
(i.e. Hessians). Because for many high-level electronic structure
methods analytical second derivatives are not explicitly implemented,
the present NN-based approach is advantageous both, in terms of
accuracy and computational efficiency. The present method can be
systematically improved by using data at higher levels of theory,
explicitly or by using TL, and allows to obtain VPT2 frequencies at
levels of theory for which \textit{ab initio} VPT2 calculation are
impractical.\\

\noindent
The current study can be extended to include IR intensities for
fundamentals, combination bands and overtones. Moreover, the approach
can be extended to investigate larger molecules and assist in
(re)assigning experimental IR spectra. It is also shown that for the
smallest molecules and the lowest level of theory considered
(MP2/aug-cc-pVTZ) direct evaluation of harmonic and VPT2 frequencies
is computationally more efficient than the combined NN + VPT2
approach. However, for molecules with 6 and more atoms and for the
higher levels of quantum chemical treatment (CCSD(T) and higher) the
NN + VPT2 approach is computationally considerably more
efficient. Finally, the VPT2 frequencies at these highest levels of
theory compare favourably (to within a few cm$^{-1}$) with
experimentally determined frequencies. One additional application for
high-level VPT2 calculations is the accurate determination of
dimerization energies, such as H-bonded complexes, for which accurate
anharmonic zero point vibrational energies are
required\cite{suhm2012dissociation}. Also, using DFT calculations for
the reference ML model in TL to higher levels of theory is an
attractive prospect to further improve the efficiency of the present
method. Finally, additional improvements may be achieved by using
sophisticated sampling or learning approaches such as active
learning.\cite{smith2018less}

\section*{Data Availability Statement}
The PhysNet codes are available at
\url{https://github.com/MMunibas/PhysNet}, and the VibML dataset
containing the reference data can be downloaded from Zenodo
\url{https://doi.org/10.5281/zenodo.4585449}.\\

\section*{Acknowledgments}
This work was supported by the Swiss National Science Foundation
through grants 200021-117810, 200020-188724 and the NCCR MUST, and the
University of Basel.

\bibliography{references}
\end{document}


\date{\today}

\section{Data sets}
\begin{table}[h]
\begin{tabular}{ccccc}
\toprule
\#&	\textbf{Molec}	&	\textbf{MP2/AVTZ}	&	\textbf{CCSD(T)/AVTZ}	&	\textbf{CCSD(T)-F12/AVTZ-F12}\\
\midrule
1&	H$_2$CO	&	0.5 min	&	5.5 min	&	18.5 min	\\
2&	HONO	&	1.0 min	&	27.0 min	&	88.0 min	\\
3&	HCOOH	&	1.5 min	&	33.5 min	&	130.0 min	\\
4&	CH$_3$OH	&	1.0 min	&	21.5 min	&	70.0 min	\\
5&	CH$_3$CHO	&	4.5 min	&	102.0 min	&	255.0 min	\\
6&	CH$_3$NO$_2$	&	5.5 min	&	268.0 min	&	876.0 min	\\
7&	CH$_3$COOH	&	9.0 min	&	380.0 min	&	??	\\
8&	CH$_3$CONH$_2$	&	11.0 min	&	600.0 min	&	??\\
\bottomrule
\end{tabular} 
\caption{CPU timings for single processor jobs for the ab initio calculations using the given levels
  of theory for a single calculation of energy, gradients and dipole
  moments with tightened convergence criteria. For entries with ??  the calculation failed either due to
  memory or space issues. Note that for example a CCSD(T)-F12 calculation for CH$_3$CHO needed about 350 GB of
  space. For molecules 1 to 3 the calculation on CCSD(T)-F12, for
  molecule 4 the calculations on CCSD(T) and for molecules 5 to 8 the calculations on the MP2 level of theory were
  performed. While the timing calculations were performed on a Intel(R) Xeon(R) CPU E5-2630 v4 @ 2.20GHz for consistency, the remaining calculations were
performed on mixed computer architectures.}\label{sitab: timings}
\end{table}

\section{Results}
\subsection{Energy- and force-errors}

\begin{sidewaystable}[h]
\resizebox{\linewidth}{!}{%
\begin{tabular}{lcccccccccccccccc}
\toprule
& \multicolumn{2}{c}{{\bf H$_2$CO}} &  \multicolumn{2}{c}{{\bf HONO}}& \multicolumn{2}{c}{{\bf HCOOH}} & \multicolumn{2}{c}{{\bf CH$_3$OH}} & \multicolumn{2}{c}{{\bf CH$_3$CHO}}& \multicolumn{2}{c}{{\bf CH$_3$COOH}}& \multicolumn{2}{c}{{\bf CH$_3$NO$_2$}}& \multicolumn{2}{c}{{\bf CH$_3$CONH$_2$}}\\
{\bf MP2} & NN$_1$ & NN$_2$ & NN$_1$ & NN$_2$ & NN$_1$ & NN$_2$ & NN$_1$ & NN$_2$ & NN$_1$ & NN$_2$ & NN$_1$ & NN$_2$ & NN$_1$ & NN$_2$& NN$_1$ & NN$_2$\\
\midrule
EMAE:	&	7.19	&	1.26	&	1.92	&	2.47	&	20.42	&	10.43	&	7.90	&	2.93	&	1.36	&	15.13	&	10.35	&	21.05	&	8.48	&	12.26	&	21.83	&	22.30	\\
ERMSE:	&	7.19	&	1.26	&	3.92	&	7.01	&	20.46	&	10.47	&	7.98	&	3.13	&	3.54	&	15.51	&	10.52	&	21.20	&	8.54	&	12.47	&	22.96	&	24.36	\\
FMAE:	&	1.42	&	1.41	&	5.79	&	5.64	&	6.75	&	6.92	&	7.44	&	7.24	&	13.25	&	13.01	&	13.50	&	14.06	&	9.65	&	8.55	&	36.49	&	40.02	\\
FRMSE:	&	3.66	&	3.44	&	23.32	&	27.01	&	29.83	&	18.33	&	27.02	&	29.08	&	53.37	&	47.07	&	38.41	&	43.00	&	73.22	&	67.00	&	106.02	&	128.84	\\
1-R2:	&	5.4E-6	&	1.7E-7	&	1.0E-8	&	3.0E-8	&	1.4E-5	&	3.6E-6	&	6.0E-7	&	9.0E-8	&	3.0E-7	&	5.8E-6	&	2.3E-6	&	9.1E-6	&	2.4E-6	&	5.2E-6	&	3.2E-6	&	3.6E-6	\\
\midrule
{\bf CCSD(T)} & NN$_1$ & NN$_2$ & NN$_1$ & NN$_2$ & NN$_1$ & NN$_2$ & NN$_1$ & NN$_2$ & NN$_1$ & NN$_2$ & NN$_1$ & NN$_2$ & NN$_1$ & NN$_2$& NN$_1$ & NN$_2$\\
\midrule																																
																																	
EMAE:	&	0.47	&	1.21	&	1.56	&	2.36	&	21.90	&	0.63	&	8.67	&	5.04	&		&		&		&		&		&		&		&		\\
ERMSE:	&	0.49	&	1.22	&	4.71	&	5.54	&	21.91	&	0.87	&	8.74	&	5.15	&		&		&		&		&		&		&		&		\\
FMAE:	&	1.48	&	1.49	&	4.87	&	5.74	&	5.93	&	6.73	&	6.56	&	7.32	&		&		&		&		&		&		&		&		\\
FRMSE:	&	4.25	&	4.19	&	14.94	&	15.21	&	21.64	&	27.07	&	22.61	&	23.59	&		&		&		&		&		&		&		&		\\
1-R2:	&	3.0E-8	&	1.6E-7	&	2.0E-8	&	2.0E-8	&	1.6E-5	&	2.0E-8	&	7.5E-7	&	2.6E-7	&		&		&		&		&		&		&		&		\\
																																	
\midrule
{\bf CCSD(T)-F12} & NN$_1$ & NN$_2$ & NN$_1$ & NN$_2$ & NN$_1$ & NN$_2$ & NN$_1$ & NN$_2$ & NN$_1$ & NN$_2$ & NN$_1$ & NN$_2$ & NN$_1$ & NN$_2$& NN$_1$ & NN$_2$\\
\midrule

EMAE:	&	0.90	&	2.44	&	6.96	&	5.90	&	3.13	&	7.60	&		&		&		&		&		&		&		&		&		&		\\
ERMSE:	&	1.00	&	2.49	&	11.27	&	6.81	&	3.25	&	7.68	&		&		&		&		&		&		&		&		&		&		\\
FMAE:	&	2.02	&	2.43	&	5.09	&	5.15	&	6.24	&	5.69	&		&		&		&		&		&		&		&		&		&		\\
FRMSE:	&	5.13	&	6.44	&	15.97	&	17.34	&	26.35	&	23.51	&		&		&		&		&		&		&		&		&		&		\\
1-R2:	&	1.1E-7	&	6.7E-7	&	9.0E-8	&	3.0E-8	&	3.5E-7	&	2.0E-6	&		&		&		&		&		&		&		&		&		&		\\

\bottomrule
\end{tabular}}
\caption{Performance measures of two PhysNet models trained independently on
the same \textit{ab initio} data and evaluated on the test set. The MAEs and RMSEs are
given in kcal/mol(/\AA) and multiplied by a factor of 1000 for clarity. }\label{sitab:all_mae}
\end{sidewaystable}

\clearpage
\subsection{H$_2$CO}
\begin{table}[]
\begin{tabular}{lrrrrr}
\toprule
Mode & NN1(MP2) &  NN2(MP2) &     MP2 &  $|\Delta 1|$ &  $|\Delta 2|$ \\
\midrule
    1 &   1196.65 &   1196.65 & 1196.70 &     0.05 &     0.05 \\
    2 &   1266.71 &   1266.70 & 1266.64 &     0.07 &     0.06 \\
    3 &   1539.99 &   1539.91 & 1539.91 &     0.08 &     0.00 \\
    4 &   1752.61 &   1752.47 & 1752.51 &     0.10 &     0.04 \\
    5 &   2973.23 &   2973.11 & 2973.27 &     0.04 &     0.16 \\
    6 &   3047.04 &   3047.10 & 3047.30 &     0.26 &     0.20 \\
\bottomrule
  \textbf{MAE:} &      0.10 &      0.09 &      &       &       \\
\end{tabular}
\caption{Normal mode frequencies (in cm$^{-1}$) of H$_2$CO calculated from PhysNet models trained on
MP2 data and compared to their reference \textit{ab initio} values. The MAEs of the
PhysNet predictions with respect to reference are given.}\label{sitab:h2co_harm_mp2}
\end{table}

\begin{table}[]
\begin{tabular}{lrrrrr}
\toprule
 Mode & NN1(CCSD(T)) &  NN2(CCSD(T)) &  CCSD(T) &  $|\Delta 1|$ &  $|\Delta 2|$ \\
\midrule
    1 &       1181.00 &       1181.02 &  1181.05 &     0.05 &     0.03 \\
    2 &       1261.67 &       1261.64 &  1261.53 &     0.14 &     0.11 \\
    3 &       1529.54 &       1529.43 &  1529.42 &     0.12 &     0.01 \\
    4 &       1764.82 &       1764.73 &  1764.87 &     0.05 &     0.14 \\
    5 &       2932.10 &       2931.96 &  2932.13 &     0.03 &     0.17 \\
    6 &       2999.94 &       2999.79 &  3000.07 &     0.13 &     0.28 \\
    \bottomrule
 \textbf{MAE:} &          0.09 &          0.13 &       &       &       \\
\end{tabular}
\caption{Normal mode frequencies (in cm$^{-1}$) of H$_2$CO calculated from PhysNet models trained on
CCSD(T) data and compared to their reference \textit{ab initio} values. The MAEs of the
PhysNet predictions with respect to reference are given.}\label{sitab:h2co_harm_cc}
\end{table}

\begin{table}[]
\begin{tabular}{lrrrrr}
\toprule
 Mode &NN1(CCSD(T)-F12) &  NN2(CCSD(T)-F12) &  CCSD(T)-F12 &  $|\Delta 1|$ &  $|\Delta 2|$ \\
\midrule
    1 &           1186.52 &           1186.50 &      1186.53 &     0.01 &     0.03 \\
    2 &           1268.15 &           1268.17 &      1268.08 &     0.07 &     0.09 \\
    3 &           1532.76 &           1532.72 &      1532.67 &     0.09 &     0.05 \\
    4 &           1776.40 &           1776.43 &      1776.53 &     0.13 &     0.10 \\
    5 &           2933.37 &           2933.57 &      2933.75 &     0.38 &     0.18 \\
    6 &           3005.45 &           3005.69 &      3005.75 &     0.30 &     0.06 \\
    \bottomrule
 \textbf{MAE:} &              0.17 &              0.08 &           &       &       \\
\end{tabular}
\caption{Normal mode frequencies (in cm$^{-1}$) of H$_2$CO calculated from PhysNet models trained on
CCSD(T)-F12 data and compared to their reference \textit{ab initio} values. The MAEs of the
PhysNet predictions with respect to reference are given.}\label{sitab:h2co_harm_ccf12}
\end{table}

\begin{sidewaystable}[h]
\resizebox{\linewidth}{!}{%
\begin{tabular}{lrrrrrrrr}
\toprule
Mode &  NN1(MP2) &  NN1(MP2) &     MP2 &  NN1(CCSD(T)) &  NN1(CCSD(T)) &  NN1(CCSD(T)-F12) &  NN1(CCSD(T)-F12) &     Exp \\
\midrule
   1 &   1180.04 &   1179.94 & 1180.23 &       1163.73 &       1163.52 &           1166.36 &           1166.73 & 1167.00 \\
   2 &   1246.72 &   1246.67 & 1246.71 &       1240.62 &       1241.36 &           1246.01 &           1245.61 & 1249.00 \\
   3 &   1507.55 &   1507.47 & 1507.95 &       1493.89 &       1494.91 &           1498.43 &           1498.36 & 1500.00 \\
   4 &   1719.02 &   1716.10 & 1720.94 &       1729.94 &       1731.88 &           1745.32 &           1744.66 & 1746.00 \\
   5 &   2823.82 &   2823.11 & 2826.67 &       2775.38 &       2779.58 &           2783.01 &           2778.06 & 2782.00 \\
   6 &   2862.29 &   2859.67 & 2862.61 &       2819.51 &       2821.49 &           2825.99 &           2821.91 & 2843.00 \\
\bottomrule
 \textbf{MAE:} &     18.49 &     18.40 &      &         10.65 &          9.04 &              3.98 &              5.28 &      \\
\end{tabular}}
\caption{VPT2 anharmonic frequencies (in cm$^{-1}$) of H$_2$CO calculated using PhysNet (NN1 and NN2)
trained on MP2, CCSD(T) and CCSD(T)-F12 data. They are compared to their reference \textit{\textit{ab initio}}
values (MP2) as well as with experiment \cite{herndon2005determination}. The MP2
frequencies are added as the VPT2 calculation is feasible in Gaussian. The MAEs are given
with respect to experiment.}\label{sitab:h2co_ah}
\end{sidewaystable}
\clearpage

\subsection{HONO}
\begin{table}[]
\begin{tabular}{lrrrrr}
\toprule
 Mode & NN1(MP2) &  NN2(MP2) &     MP2 &  $|\Delta 1|$ &  $|\Delta 2|$ \\
\midrule
    1 &    586.44 &    586.45 &  586.53 &     0.09 &     0.08 \\
    2 &    602.45 &    602.40 &  602.38 &     0.07 &     0.02 \\
    3 &    805.28 &    805.18 &  805.20 &     0.08 &     0.02 \\
    4 &   1283.39 &   1283.37 & 1283.28 &     0.11 &     0.09 \\
    5 &   1659.77 &   1659.76 & 1659.81 &     0.04 &     0.05 \\
    6 &   3754.55 &   3754.40 & 3754.95 &     0.40 &     0.55 \\
    \bottomrule
 \textbf{MAE:} &      0.13 &      0.13 &      &       &       \\

\end{tabular}
\caption{Normal mode frequencies (in cm$^{-1}$) of HONO calculated from PhysNet models trained on
MP2 data and compared to their reference \textit{ab initio} values. The MAEs of the
PhysNet predictions with respect to reference are given.}\label{sitab:hono_harm_mp2}
\end{table}

\begin{table}[]
\begin{tabular}{lrrrrr}
\toprule
Mode & NN1(CCSD(T)) &  NN2(CCSD(T)) &     CCSD(T) &  $|\Delta 1|$ &  $|\Delta 2|$ \\
\midrule
    1 &        565.06 &        565.02 &   564.84 &     0.22 &     0.18 \\
    2 &        617.31 &        617.34 &   617.20 &     0.11 &     0.14 \\
    3 &        815.84 &        815.85 &   815.79 &     0.05 &     0.06 \\
    4 &       1305.94 &       1305.95 &  1305.94 &     0.00 &     0.01 \\
    5 &       1715.12 &       1715.15 &  1715.35 &     0.23 &     0.20 \\
    6 &       3759.58 &       3759.68 &  3760.30 &     0.72 &     0.62 \\
\bottomrule
 \textbf{MAE:} &          0.22 &          0.20 &       &       &       \\
\end{tabular}
\caption{Normal mode frequencies (in cm$^{-1}$) of HONO calculated from PhysNet models trained on
CCSD(T) data and compared to their reference \textit{ab initio} values. The MAEs of the
PhysNet predictions with respect to reference are given.}\label{sitab:hono_harm_cc}
\end{table}

\begin{table}[]
\begin{tabular}{lrrrrr}
\toprule
 Mode & NN1(CCSD(T)-F12) &  NN2(CCSD(T)-F12) &  CCSD(T)-F12 &  $|\Delta 1|$ &  $|\Delta 2|$ \\
\midrule
    1 &            573.12 &            573.07 &       573.02 &     0.10 &     0.05 \\
    2 &            631.54 &            631.55 &       631.40 &     0.14 &     0.15 \\
    3 &            831.14 &            831.23 &       831.14 &     0.00 &     0.09 \\
    4 &           1315.91 &           1315.85 &      1315.76 &     0.15 &     0.09 \\
    5 &           1733.36 &           1733.39 &      1733.55 &     0.19 &     0.16 \\
    6 &           3774.84 &           3774.96 &      3775.54 &     0.70 &     0.58 \\
\bottomrule
 \textbf{MAE:} &              0.21 &              0.19 &           &       &       \\
\end{tabular}
\caption{Normal mode frequencies (in cm$^{-1}$) of HONO calculated from PhysNet models trained on
CCSD(T)-F12 data and compared to their reference \textit{ab initio} values. The MAEs of the
PhysNet predictions with respect to reference are given.}\label{sitab:hono_harm_ccf12}
\end{table}
\clearpage
\begin{sidewaystable}[h]
\resizebox{\linewidth}{!}{%
\begin{tabular}{lrrrrrrrr}
\toprule
Mode &  NN1(MP2) &  NN1(MP2) &     MP2 &  NN1(CCSD(T)) &  NN1(CCSD(T)) &  NN1(CCSD(T)-F12) &  NN1(CCSD(T)-F12) &     Exp \\
\midrule
   1 &    551.59 &    551.72 &  551.35 &        533.67 &        533.93 &            541.10 &            542.37 &  543.88 \\
   2 &    566.02 &    566.12 &  565.97 &        595.98 &        596.32 &            608.55 &            608.92 &  595.62 \\
   3 &    766.25 &    765.86 &  765.83 &        785.91 &        786.93 &            799.70 &            799.04 &  790.12 \\
   4 &   1235.12 &   1230.53 & 1233.45 &       1260.08 &       1260.91 &           1267.74 &           1267.88 & 1263.21 \\
   5 &   1637.07 &   1636.84 & 1637.31 &       1691.60 &       1691.48 &           1709.88 &           1710.24 & 1699.76 \\
   6 &   3580.86 &   3581.00 & 3577.17 &       3580.01 &       3581.84 &           3590.28 &           3586.80 & 3590.77 \\
\bottomrule
 \textbf{MAE:} &     26.98 &     27.83 &      &          6.14 &          5.56 &              6.74 &              7.14 &      \\
\end{tabular}}
\caption{VPT2 anharmonic frequencies (in cm$^{-1}$) of HONO calculated using PhysNet (NN1 and NN2)
trained on MP2, CCSD(T) and CCSD(T)-F12 data. They are compared to their reference \textit{\textit{ab initio}}
values (MP2) as well as with experiment \cite{guilmot1993rovibrational}. The MP2
frequencies are added as the VPT2 calculation is feasible in Gaussian. The MAEs are given
with respect to experiment.}\label{sitab:hono_ah}
\end{sidewaystable}
\clearpage

\subsection{HCOOH}
\begin{table}[]
\begin{tabular}{lrrrrr}
\toprule
Mode & NN1(MP2) &  NN2(MP2) &     MP2 &  $|\Delta 1|$ &  $|\Delta 2|$ \\
\midrule
    1 &    625.93 &    625.97 &  625.86 &     0.07 &     0.11 \\
    2 &    675.16 &    675.14 &  675.28 &     0.12 &     0.14 \\
    3 &   1058.72 &   1058.68 & 1058.77 &     0.05 &     0.09 \\
    4 &   1130.58 &   1130.91 & 1130.60 &     0.02 &     0.31 \\
    5 &   1301.64 &   1301.70 & 1301.56 &     0.08 &     0.14 \\
    6 &   1409.12 &   1409.07 & 1408.91 &     0.21 &     0.16 \\
    7 &   1793.39 &   1793.17 & 1793.29 &     0.10 &     0.12 \\
    8 &   3123.89 &   3123.92 & 3123.94 &     0.05 &     0.02 \\
    9 &   3740.49 &   3740.31 & 3740.56 &     0.07 &     0.25 \\
\bottomrule
 \textbf{MAE:} &      0.08 &      0.15 &      &       &       \\

\end{tabular}
\caption{Normal mode frequencies (in cm$^{-1}$) of HCOOH calculated from PhysNet models trained on
MP2 data and compared to their reference \textit{ab initio} values. The MAEs of the
PhysNet predictions with respect to reference are given.}\label{sitab:hcooh_harm_mp2}
\end{table}

\begin{table}[]
\begin{tabular}{lrrrrr}
\toprule
Mode & NN1(CCSD(T)) &  NN2(CCSD(T)) &  CCSD(T) &  $|\Delta 1|$ &  $|\Delta 2|$ \\
\midrule
    1 &        626.52 &        626.68 &   626.46 &     0.06 &     0.22 \\
    2 &        664.46 &        664.43 &   664.94 &     0.48 &     0.51 \\
    3 &       1050.94 &       1050.93 &  1050.99 &     0.05 &     0.06 \\
    4 &       1131.48 &       1131.52 &  1131.27 &     0.21 &     0.25 \\
    5 &       1310.97 &       1311.12 &  1310.78 &     0.19 &     0.34 \\
    6 &       1404.92 &       1404.91 &  1404.71 &     0.21 &     0.20 \\
    7 &       1802.52 &       1802.56 &  1802.63 &     0.11 &     0.07 \\
    8 &       3088.01 &       3087.86 &  3087.61 &     0.40 &     0.25 \\
    9 &       3741.66 &       3741.54 &  3741.84 &     0.18 &     0.30 \\
\bottomrule
    \textbf{MAE:} &          0.21 &          0.24 &       &       &       \\
\end{tabular}
\caption{Normal mode frequencies (in cm$^{-1}$) of HCOOH calculated from PhysNet models trained on
CCSD(T) data and compared to their reference \textit{ab initio} values. The MAEs of the
PhysNet predictions with respect to reference are given.}\label{sitab:hcooh_harm_cc}
\end{table}

\begin{table}[]
\begin{tabular}{lrrrrr}
\toprule
Mode & NN1(CCSD(T)-F12) &  NN2(CCSD(T)-F12) &  CCSD(T)-F12 &  $|\Delta 1|$ &  $|\Delta 2|$ \\
\midrule
    1 &            630.68 &            630.66 &       630.56 &     0.12 &     0.10 \\
    2 &            672.06 &            672.06 &       672.14 &     0.08 &     0.08 \\
    3 &           1055.17 &           1055.17 &      1055.23 &     0.06 &     0.06 \\
    4 &           1136.51 &           1136.47 &      1136.26 &     0.25 &     0.21 \\
    5 &           1315.68 &           1315.68 &      1315.48 &     0.20 &     0.20 \\
    6 &           1407.20 &           1407.08 &      1406.99 &     0.21 &     0.09 \\
    7 &           1811.85 &           1811.79 &      1811.81 &     0.04 &     0.02 \\
    8 &           3092.10 &           3091.79 &      3091.90 &     0.20 &     0.11 \\
    9 &           3755.50 &           3755.53 &      3755.69 &     0.19 &     0.16 \\
\bottomrule
 \textbf{MAE:} &              0.15 &              0.11 &           &       &       \\
\end{tabular}
\caption{Normal mode frequencies (in cm$^{-1}$) of HCOOH calculated from PhysNet models trained on
CCSD(T)-F12 data and compared to their reference \textit{ab initio} values. The MAEs of the
PhysNet predictions with respect to reference are given.}\label{sitab:hcooh_harm_ccf12}
\end{table}

\clearpage
\begin{sidewaystable}[h]
\resizebox{\linewidth}{!}{%
\begin{tabular}{lrrrrrrrr}
\toprule
Mode &  NN1(MP2) &  NN1(MP2) &     MP2 &  NN1(CCSD(T)) &  NN1(CCSD(T)) &  NN1(CCSD(T)-F12) &  NN1(CCSD(T)-F12) &     Exp \\
\midrule
   1 &    619.28 &    618.68 &  619.54 &        620.37 &        619.30 &            626.05 &            625.78 &  626.16 \\
   2 &    642.69 &    641.62 &  641.78 &        630.86 &        630.52 &            637.65 &            637.58 &  640.72 \\
   3 &   1037.06 &   1036.44 & 1036.38 &       1028.34 &       1028.38 &           1033.08 &           1033.04 & 1033.47 \\
   4 &   1097.01 &   1096.67 & 1098.18 &       1099.54 &       1098.65 &           1104.11 &           1104.70 & 1104.85 \\
   5 &   1220.49 &   1219.25 & 1220.63 &       1294.92 &       1293.36 &           1301.99 &           1301.89 & 1306.20 \\
   6 &   1380.81 &   1380.24 & 1380.94 &       1375.30 &       1374.86 &           1377.35 &           1378.23 & 1380.00 \\
   7 &   1758.17 &   1757.89 & 1760.53 &       1765.98 &       1768.89 &           1774.44 &           1774.60 & 1776.83 \\
   8 &   2959.50 &   2960.15 & 2967.80 &       2923.05 &       2922.24 &           2924.61 &           2927.30 & 2942.00 \\
   9 &   3554.88 &   3555.83 & 3554.90 &       3557.15 &       3562.98 &           3565.70 &           3565.34 & 3570.50 \\
\bottomrule
 \textbf{MAE:} &     17.62 &     17.61 &      &          9.47 &          9.06 &              3.97 &              3.59 &      \\
\end{tabular}}
\caption{VPT2 anharmonic frequencies (in cm$^{-1}$) of HCOOH calculated using PhysNet (NN1 and NN2)
trained on MP2, CCSD(T) and CCSD(T)-F12 data. They are compared to their reference \textit{\textit{ab initio}}
values (MP2) as well as with experiment \cite{tew2016ab}. The MP2
frequencies are added as the VPT2 calculation is feasible in Gaussian. The MAEs are given
with respect to experiment.}\label{sitab:hcooh_ah}
\end{sidewaystable}
\clearpage

\clearpage
\subsection{CH$_3$OH}
\begin{table}[]
\begin{tabular}{lrrrrr}
\toprule
Mode& NN1(MP2) &  NN2(MP2) &     MP2 &  $|\Delta 1|$ &  $|\Delta 2|$ \\
\midrule
    1 &    289.86 &    289.85 &  289.38 &     0.48 &     0.47 \\
    2 &   1056.62 &   1056.61 & 1056.64 &     0.02 &     0.03 \\
    3 &   1088.65 &   1088.59 & 1088.63 &     0.02 &     0.04 \\
    4 &   1183.93 &   1183.91 & 1183.87 &     0.06 &     0.04 \\
    5 &   1371.59 &   1371.60 & 1371.57 &     0.02 &     0.03 \\
    6 &   1491.97 &   1491.96 & 1492.01 &     0.04 &     0.05 \\
    7 &   1525.40 &   1525.42 & 1525.41 &     0.01 &     0.01 \\
    8 &   1535.71 &   1535.79 & 1535.80 &     0.09 &     0.01 \\
    9 &   3053.97 &   3054.05 & 3053.94 &     0.03 &     0.11 \\
   10 &   3125.24 &   3125.24 & 3125.08 &     0.16 &     0.16 \\
   11 &   3182.86 &   3182.88 & 3182.91 &     0.05 &     0.03 \\
   12 &   3859.73 &   3859.80 & 3859.62 &     0.11 &     0.18 \\
\bottomrule
 \textbf{MAE:} &      0.09 &      0.10 &      &       &       \\
\end{tabular}
\caption{Normal mode frequencies (in cm$^{-1}$) of CH$_3$OH calculated from PhysNet models trained on
MP2 data and compared to their reference \textit{ab initio} values. The MAEs of the
PhysNet predictions with respect to reference are given.}\label{sitab:ch3oh_harm_mp2}
\end{table}

\begin{table}[]
\begin{tabular}{lrrrrr}
\toprule
Mode & NN1(CCSD(T)) &  NN2(CCSD(T)) &  CCSD(T)& $|\Delta 1|$ &  $|\Delta 2|$ \\
\midrule
    1 &        286.35 &        286.33 &   285.95 &     0.40 &     0.38 \\
    2 &       1053.43 &       1053.41 &  1053.38 &     0.05 &     0.03 \\
    3 &       1082.20 &       1082.18 &  1082.11 &     0.09 &     0.07 \\
    4 &       1175.88 &       1175.85 &  1175.72 &     0.16 &     0.13 \\
    5 &       1379.10 &       1379.11 &  1379.01 &     0.09 &     0.10 \\
    6 &       1483.92 &       1483.91 &  1483.86 &     0.06 &     0.05 \\
    7 &       1512.18 &       1512.20 &  1512.16 &     0.02 &     0.04 \\
    8 &       1522.63 &       1522.65 &  1522.67 &     0.04 &     0.02 \\
    9 &       3010.68 &       3010.64 &  3010.43 &     0.25 &     0.21 \\
   10 &       3068.92 &       3068.88 &  3068.65 &     0.27 &     0.23 \\
   11 &       3128.02 &       3127.96 &  3127.97 &     0.05 &     0.01 \\
   12 &       3843.44 &       3843.41 &  3843.27 &     0.17 &     0.14 \\
\bottomrule
 \textbf{MAE:} &          0.14 &          0.12 &       &       &       \\
\end{tabular}
\caption{Normal mode frequencies (in cm$^{-1}$) of CH$_3$OH calculated from PhysNet models trained on
CCSD(T) data and compared to their reference \textit{ab initio} values. The MAEs of the
PhysNet predictions with respect to reference are given.}\label{sitab:ch3oh_harm_cc}
\end{table}

\clearpage
\begin{table}[]
\begin{tabular}{lrrrrrr}
\toprule
Mode &  NN1(MP2) &  NN1(MP2) &     MP2 &  NN1(CCSD(T)) &  NN1(CCSD(T)) &     Exp \\
\midrule
   1 &    240.84 &    241.49 &  240.64 &        236.97 &        238.08 &  271.50 \\
   2 &   1031.16 &   1030.51 & 1030.29 &       1026.41 &       1027.04 & 1033.50 \\
   3 &   1069.23 &   1069.65 & 1069.25 &       1063.32 &       1063.78 & 1074.50 \\
   4 &   1154.53 &   1154.90 & 1155.02 &       1145.68 &       1145.95 & 1145.00 \\
   5 &   1321.89 &   1321.99 & 1321.98 &       1321.31 &       1322.05 & 1332.00 \\
   6 &   1457.42 &   1457.78 & 1457.30 &       1447.89 &       1448.52 & 1454.50 \\
   7 &   1483.39 &   1483.11 & 1483.22 &       1469.21 &       1468.55 & 1465.00 \\
   8 &   1490.11 &   1489.77 & 1489.48 &       1475.75 &       1475.62 & 1479.50 \\
   9 &   2998.34 &   2997.26 & 2996.01 &       2827.05 &       2829.22 & 2844.20 \\
  10 &   3000.01 &   3001.00 & 2996.40 &       2911.42 &       2912.12 & 2970.00 \\
  11 &   3048.63 &   3048.07 & 3046.42 &       2987.57 &       2988.73 & 2999.00 \\
  12 &   3688.76 &   3686.33 & 3687.39 &       3666.53 &       3667.99 & 3681.50 \\
\bottomrule
 \textbf{MAE:} &     27.57 &     27.28 &      &         15.07 &         14.30 &      \\
\end{tabular}
\caption{VPT2 anharmonic frequencies (in cm$^{-1}$) of CH$_3$OH calculated using PhysNet (NN1 and NN2)
trained on MP2 and CCSD(T) data. They are compared to their reference \textit{\textit{ab initio}}
values (MP2) as well as with experiment \cite{serrallach1974methanol}. The MP2
frequencies are added as the VPT2 calculation is feasible in Gaussian. The MAEs are given
with respect to experiment. Note that the lowest experimental
frequency was obtained from Ar-matrix measurements.}\label{sitab:ch3oh_ah}
\end{table}


\clearpage
\subsection{CH$_3$CHO}
\begin{table}[]
\begin{tabular}{lrrrrr}
\toprule
Mode & NN1(MP2) &  NN2(MP2) &     MP2 &  $|\Delta 1|$ &  $|\Delta 2|$ \\
\midrule
    1 &    160.92 &    160.91 &  160.54 &     0.38 &     0.37 \\
    2 &    506.62 &    506.67 &  506.58 &     0.04 &     0.09 \\
    3 &    779.28 &    779.25 &  779.24 &     0.04 &     0.01 \\
    4 &    905.34 &    905.34 &  905.34 &     0.00 &     0.00 \\
    5 &   1136.34 &   1136.34 & 1136.37 &     0.03 &     0.03 \\
    6 &   1143.70 &   1143.72 & 1143.64 &     0.06 &     0.08 \\
    7 &   1390.46 &   1390.48 & 1390.38 &     0.08 &     0.10 \\
    8 &   1426.61 &   1426.58 & 1426.75 &     0.14 &     0.17 \\
    9 &   1481.35 &   1481.32 & 1481.20 &     0.15 &     0.12 \\
   10 &   1493.68 &   1493.73 & 1493.54 &     0.14 &     0.19 \\
   11 &   1763.36 &   1763.45 & 1763.86 &     0.50 &     0.41 \\
   12 &   2955.29 &   2955.30 & 2954.41 &     0.88 &     0.89 \\
   13 &   3069.58 &   3069.48 & 3069.68 &     0.10 &     0.20 \\
   14 &   3149.19 &   3149.18 & 3149.53 &     0.34 &     0.35 \\
   15 &   3199.78 &   3199.83 & 3199.44 &     0.34 &     0.39 \\
\bottomrule
 \textbf{MAE:} &      0.22 &      0.23 &      &       &       \\
\end{tabular}
\caption{Normal mode frequencies (in cm$^{-1}$) of CH$_3$CHO calculated from PhysNet models trained on
MP2 data and compared to their reference \textit{ab initio} values. The MAEs of the
PhysNet predictions with respect to reference are given.}\label{sitab:ch3cho_harm_mp2}
\end{table}

\clearpage
\begin{table}[]
\begin{tabular}{lrrrr}
\toprule
Mode &  NN1(MP2) &  NN1(MP2) &     MP2 &     Exp \\
\midrule
   1 &    150.37 &    151.46 &  149.94 &  143.80 \\
   2 &    505.60 &    505.45 &  505.96 &  508.80 \\
   3 &    766.03 &    766.42 &  765.11 &  764.10 \\
   4 &    867.46 &    868.57 &  867.05 &  865.90 \\
   5 &   1110.82 &   1110.70 & 1109.98 & 1097.80 \\
   6 &   1115.91 &   1116.88 & 1116.41 & 1113.80 \\
   7 &   1352.20 &   1351.93 & 1351.80 & 1352.60 \\
   8 &   1403.31 &   1403.48 & 1403.23 & 1394.90 \\
   9 &   1437.73 &   1438.49 & 1437.76 & 1433.50 \\
  10 &   1445.30 &   1444.24 & 1445.08 & 1436.30 \\
  11 &   1728.74 &   1726.97 & 1726.61 & 1746.00 \\
  12 &   2735.11 &   2735.32 & 2735.51 & 2715.40 \\
  13 &   2966.08 &   2965.81 & 2964.93 & 2923.20 \\
  14 &   3012.46 &   3011.77 & 3012.12 & 2964.30 \\
  15 &   3064.91 &   3060.87 & 3059.86 & 3014.30 \\
\bottomrule
 \textbf{MAE:} &     15.27 &     15.32 &      &      \\
\end{tabular}
\caption{VPT2 anharmonic frequencies (in cm$^{-1}$) of CH$_3$CHO calculated using PhysNet (NN1 and NN2)
trained on MP2 data. They are compared to their reference \textit{\textit{ab initio}}
values (MP2) as well as with experiment \cite{wiberg1995acetaldehyde}. The MP2
frequencies are added as the VPT2 calculation is feasible in Gaussian. The MAEs are given
with respect to experiment.}\label{sitab:ch3cho_ah}
\end{table}
\clearpage

\clearpage
\subsection{CH$_3$COOH}
\begin{table}[]
\begin{tabular}{lrrrrr}
\toprule
Mode & NN1(MP2) &  NN2(MP2) &     MP2 &  $|\Delta 1|$ &  $|\Delta 2|$ \\
\midrule
    1 &     77.46 &     77.49 &   77.20 &     0.26 &     0.29 \\
    2 &    422.71 &    422.66 &  422.69 &     0.02 &     0.03 \\
    3 &    549.67 &    549.67 &  549.71 &     0.04 &     0.04 \\
    4 &    583.21 &    583.18 &  583.15 &     0.06 &     0.03 \\
    5 &    661.91 &    661.90 &  662.07 &     0.16 &     0.17 \\
    6 &    873.85 &    873.76 &  873.74 &     0.11 &     0.02 \\
    7 &   1008.24 &   1008.20 & 1008.18 &     0.06 &     0.02 \\
    8 &   1075.99 &   1075.94 & 1075.86 &     0.13 &     0.08 \\
    9 &   1206.09 &   1205.93 & 1205.89 &     0.20 &     0.04 \\
   10 &   1342.18 &   1342.04 & 1342.07 &     0.11 &     0.03 \\
   11 &   1422.08 &   1421.89 & 1421.91 &     0.17 &     0.02 \\
   12 &   1492.58 &   1492.45 & 1492.36 &     0.22 &     0.09 \\
   13 &   1501.13 &   1500.97 & 1500.98 &     0.15 &     0.01 \\
   14 &   1809.90 &   1809.91 & 1810.19 &     0.29 &     0.28 \\
   15 &   3097.30 &   3097.21 & 3097.34 &     0.04 &     0.13 \\
   16 &   3180.81 &   3180.82 & 3181.01 &     0.20 &     0.19 \\
   17 &   3223.21 &   3223.25 & 3223.17 &     0.04 &     0.08 \\
   18 &   3751.61 &   3751.59 & 3751.57 &     0.04 &     0.02 \\
\bottomrule
 \textbf{MAE:} &      0.13 &      0.09 &      &       &       \\
\end{tabular}
\caption{Normal mode frequencies (in cm$^{-1}$) of CH$_3$COOH calculated from PhysNet models trained on
MP2 data and compared to their reference \textit{ab initio} values. The MAEs of the
PhysNet predictions with respect to reference are given.}\label{sitab:ch3cooh_harm_mp2}
\end{table}

\clearpage
\begin{table}[]
\begin{tabular}{lrrrr}
\toprule
Mode &  NN1(MP2) &  NN1(MP2) &     MP2 &     Exp \\
\midrule
   1 &     74.53 &     73.11 &   77.93 &    -- \\
   2 &    422.52 &    423.21 &  422.94 &  424.00 \\
   3 &    536.36 &    538.21 &  536.39 &  534.50 \\
   4 &    574.91 &    575.17 &  575.61 &  581.50 \\
   5 &    638.93 &    641.25 &  639.50 &  642.00 \\
   6 &    855.20 &    855.37 &  855.10 &  847.00 \\
   7 &    987.08 &    987.15 &  987.46 &  991.00 \\
   8 &   1047.38 &   1045.69 & 1047.31 & 1049.00 \\
   9 &   1159.13 &   1161.15 & 1159.17 & 1184.00 \\
  10 &   1320.53 &   1322.89 & 1320.83 & 1266.00 \\
  11 &   1377.70 &   1378.64 & 1378.53 & 1384.50 \\
  12 &   1438.15 &   1437.89 & 1439.69 & 1430.00 \\
  13 &   1449.95 &   1450.01 & 1450.29 & 1430.00 \\
  14 &   1780.63 &   1787.68 & 1780.90 & 1792.00 \\
  15 &   2984.12 &   2989.62 & 2989.40 & 2944.00 \\
  16 &   3037.04 &   3040.77 & 3040.41 & 2996.00 \\
  17 &   3082.90 &   3080.65 & 3080.19 & 3051.00 \\
  18 &   3574.65 &   3567.07 & 3568.62 & 3585.50 \\
\bottomrule
 \textbf{MAE:} &     16.25 &     16.67 &      &      \\
\end{tabular}
\caption{VPT2 anharmonic frequencies (in cm$^{-1}$) of CH$_3$COOH calculated using PhysNet (NN1 and NN2)
trained on MP2 data. They are compared to their reference \textit{\textit{ab initio}}
values (MP2) as well as with experiment \cite{goubet2015standard}. The MP2
frequencies are added as the VPT2 calculation is feasible in Gaussian. The MAEs are given
with respect to experiment. The lowest frequency was
not reported in Ref.~\citenum{goubet2015standard}.}\label{sitab:ch3cooh_ah}
\end{table}
\clearpage

\newpage
\subsection{CH$_3$NO$_2$}

\begin{table}[]
\begin{tabular}{lrrrrr}
\toprule
 Mode & NN1(MP2) &  NN2(MP2) &     MP2 &  $|\Delta 1|$ &  $|\Delta 2|$ \\
\midrule
    1 &     27.68 &     27.54 &   28.91 &     1.23 &     1.37 \\
    2 &    478.64 &    478.70 &  478.65 &     0.01 &     0.05 \\
    3 &    610.36 &    610.29 &  610.43 &     0.07 &     0.14 \\
    4 &    669.74 &    669.73 &  669.67 &     0.07 &     0.06 \\
    5 &    940.48 &    940.47 &  940.48 &     0.00 &     0.01 \\
    6 &   1127.26 &   1127.31 & 1127.28 &     0.02 &     0.03 \\
    7 &   1148.84 &   1148.94 & 1148.99 &     0.15 &     0.05 \\
    8 &   1411.99 &   1411.96 & 1412.12 &     0.13 &     0.16 \\
    9 &   1430.56 &   1430.56 & 1430.54 &     0.02 &     0.02 \\
   10 &   1491.99 &   1492.04 & 1491.90 &     0.09 &     0.14 \\
   11 &   1502.66 &   1502.71 & 1502.67 &     0.01 &     0.04 \\
   12 &   1745.37 &   1745.15 & 1745.72 &     0.35 &     0.57 \\
   13 &   3115.32 &   3115.35 & 3115.24 &     0.08 &     0.11 \\
   14 &   3221.42 &   3221.49 & 3221.29 &     0.13 &     0.20 \\
   15 &   3247.85 &   3247.85 & 3247.61 &     0.24 &     0.24 \\
\bottomrule
  \textbf{MAE:} &      0.17 &      0.21 &      &       &       \\
  \end{tabular}
\caption{Normal mode frequencies (in cm$^{-1}$) of CH$_3$NO$_2$ calculated from PhysNet models trained on
MP2 data and compared to their reference \textit{ab initio} values. The MAEs of the
PhysNet predictions with respect to reference are given.}\label{sitab:ch3no2_harm_mp2}
\end{table}

\clearpage
\begin{table}[]
\begin{tabular}{lrrrr}
\toprule
Mode &  NN1(MP2) &  NN1(MP2) &     MP2 &     Exp \\
\midrule
   1 &    -84.94 &    -55.88 &  -47.12 &   -- \\
   2 &    479.22 &    478.01 &  477.77 &  479.00 \\
   3 &    593.12 &    594.60 &  593.54 &  599.00 \\
   4 &    661.89 &    662.60 &  663.10 &  647.00 \\
   5 &    922.30 &    925.33 &  923.87 &  921.00 \\
   6 &   1104.10 &   1106.03 & 1105.09 & 1097.00 \\
   7 &   1120.33 &   1119.07 & 1119.84 & 1153.00 \\
   8 &   1385.86 &   1385.15 & 1385.96 & 1384.00 \\
   9 &   1394.18 &   1394.87 & 1395.09 & 1413.00 \\
  10 &   1447.35 &   1448.15 & 1448.14 & 1449.00 \\
  11 &   1448.21 &   1450.30 & 1449.79 & 1488.00 \\
  12 &   1713.45 &   1726.72 & 1727.56 & 1582.00 \\
  13 &   3011.56 &   3008.23 & 3006.22 & 2965.00 \\
  14 &   3086.54 &   3087.68 & 3086.01 & 3048.00 \\
  15 &   3107.14 &   3101.98 & 3106.89 & 3048.00 \\
  \bottomrule
 \textbf{MAE:} &     28.56 &     29.12 &      &      \\
\end{tabular}
\caption{VPT2 anharmonic frequencies (in cm$^{-1}$) of CH$_3$NO$_2$ calculated using PhysNet (NN1 and NN2)
trained on MP2 data. They are compared to their reference \textit{\textit{ab initio}}
values (MP2) as well as with experiment \cite{wells1941infra}. The MP2
frequencies are added as the VPT2 calculation is feasible in Gaussian. The MAEs are given
with respect to experiment.  The lowest frequency was
not reported in Ref.~\citenum{wells1941infra} and was negative in the VPT2 calculation.}\label{sitab:ch3no2_ah_mp2}
\end{table}
\clearpage

\subsection{CH$_3$CONH$_2$}
\begin{table}[]
\begin{tabular}{lrrrrr}
\toprule
Mode & NN1(MP2) &  NN2(MP2) &     MP2 &  $\Delta 1$ &  $\Delta 2$ \\
\midrule
    1 &     33.70 &     31.89 &   33.19 &     0.51 &     1.30 \\
    2 &    137.63 &    134.99 &  139.64 &     2.01 &     4.65 \\
    3 &    427.85 &    428.12 &  427.97 &     0.12 &     0.15 \\
    4 &    521.86 &    522.31 &  522.32 &     0.46 &     0.01 \\
    5 &    547.79 &    547.63 &  547.83 &     0.04 &     0.20 \\
    6 &    659.90 &    660.50 &  660.40 &     0.50 &     0.10 \\
    7 &    861.13 &    861.06 &  861.05 &     0.08 &     0.01 \\
    8 &    989.55 &    989.63 &  989.39 &     0.16 &     0.24 \\
    9 &   1060.85 &   1060.61 & 1060.44 &     0.41 &     0.17 \\
   10 &   1119.95 &   1119.96 & 1119.90 &     0.05 &     0.06 \\
   11 &   1353.38 &   1353.30 & 1353.21 &     0.17 &     0.09 \\
   12 &   1412.89 &   1412.92 & 1412.90 &     0.01 &     0.02 \\
   13 &   1492.30 &   1492.32 & 1492.32 &     0.02 &     0.00 \\
   14 &   1510.95 &   1510.80 & 1510.96 &     0.01 &     0.16 \\
   15 &   1623.80 &   1624.11 & 1623.89 &     0.09 &     0.22 \\
   16 &   1765.77 &   1765.60 & 1765.94 &     0.17 &     0.34 \\
   17 &   3086.67 &   3086.85 & 3086.75 &     0.08 &     0.10 \\
   18 &   3174.52 &   3174.68 & 3174.53 &     0.01 &     0.15 \\
   19 &   3198.69 &   3198.06 & 3198.43 &     0.26 &     0.37 \\
   20 &   3619.17 &   3619.10 & 3618.80 &     0.37 &     0.30 \\
   21 &   3769.75 &   3769.86 & 3769.35 &     0.40 &     0.51 \\
   \bottomrule
  \textbf{MAE:} &      0.28 &      0.44 &      &       &       \\
\end{tabular}
\caption{Normal mode frequencies (in cm$^{-1}$) of CH$_3$CONH$_2$ calculated from PhysNet models trained on
MP2 data and compared to their reference \textit{ab initio} values. The MAEs of the
PhysNet predictions with respect to reference are given.}\label{sitab:ch3conh2_harm_mp2}
\end{table}

\clearpage
\begin{table}[]
\begin{tabular}{lrrrr}
\toprule
Mode &  NN1(MP2) &  NN1(MP2) &      MP2 &     Exp \\
\midrule
   1 &  -2337.85 &  -2548.69 & -2425.72 &    -- \\
   2 &   -206.36 &   -204.12 &  -213.48 &  269.00 \\
   3 &    453.25 &    457.15 &   451.29 &  427.00 \\
   4 &    479.72 &    473.40 &   477.92 &  507.00 \\
   5 &    574.42 &    576.91 &   574.20 &  548.00 \\
   6 &    701.19 &    706.62 &   700.92 &  625.00 \\
   7 &    839.33 &    840.76 &   840.39 &  858.00 \\
   8 &    981.95 &    984.00 &   979.28 &  965.00 \\
   9 &   1037.24 &   1038.03 &  1033.17 & 1040.00 \\
  10 &   1056.99 &   1054.48 &  1054.59 & 1134.00 \\
  11 &   1342.26 &   1344.85 &  1340.62 & 1319.00 \\
  12 &   1370.49 &   1370.80 &  1371.24 & 1385.00 \\
  13 &   1444.48 &   1443.41 &  1443.50 & 1432.00 \\
  14 &   1464.30 &   1459.85 &  1463.58 & 1433.00 \\
  15 &   1556.28 &   1556.57 &  1552.95 & 1600.00 \\
  16 &   1718.97 &   1726.61 &  1725.92 & 1733.00 \\
  17 &   2996.35 &   2990.34 &  2994.98 & 2860.00 \\
  18 &   3005.78 &   2993.60 &  3008.41 & 2900.00 \\
  19 &   3068.72 &   3072.59 &  3070.81 & 2967.00 \\
  20 &   3496.12 &   3504.90 &  3501.75 & 3450.00 \\
  21 &   3646.50 &   3665.08 &  3657.46 & 3550.00 \\
\bottomrule
 \textbf{MAE:} &    47.23 &    48.40 &       &      \\
\end{tabular}
\caption{VPT2 anharmonic frequencies (in cm$^{-1}$) of CH$_3$CONH$_2$ calculated using PhysNet (NN1 and NN2)
trained on MP2 data. They are compared to their reference \textit{\textit{ab initio}}
values (MP2) as well as with experiment \cite{ganeshsrinivas1996simulation}. The MP2
frequenices are added as the VPT2 calculation is feasible in Gaussian.
Only 20 modes are assigned.
Note that four frequencies were obtained in argon matrix (269, 427, 1432, 1433).}\label{sitab:ch3conh2_ah_mp2}
\end{table}
\clearpage

\section{Transfer learning}
\subsection{H$_2$CO}
\begin{table}[]
\begin{tabular}{rrr}
\toprule
NN(TL)  & CCSD(T)-F12 & $|\Delta|$\\
\midrule
          1186.45 &      1186.53 &       0.08 \\
          1268.20 &      1268.08 &       0.12 \\
          1532.73 &      1532.67 &       0.06 \\
          1776.55 &      1776.53 &       0.02 \\
          2934.04 &      2933.75 &       0.29 \\
          3006.04 &      3005.75 &       0.29 \\
\bottomrule
\end{tabular}
\caption{Normal mode frequencies of H$_2$CO calculated from a TL model trained on
CCSD(T)-F12 data and compared to their reference \textit{ab initio} values.
All frequencies are given in cm$^{-1}$. }\label{sitab:tl_h2co_ccf12_harm}
\end{table}

\begin{table}[]
\begin{tabular}{llll}
\toprule
 NN(TL) &     Exp &  $|\Delta|$ \\
\midrule
  1167.39 & 1167.00 &       0.39 \\
  1244.93 & 1249.00 &       4.07 \\
  1497.82 & 1500.00 &       2.18 \\
  1745.22 & 1746.00 &       0.78 \\
  2770.34 & 2782.00 &      11.66 \\
  2820.39 & 2843.00 &      22.61 \\
\bottomrule
\end{tabular}
\caption{VPT2 anharmonic frequencies of H$_2$CO calculated using a TL PhysNet model
trained on CCSD(T)-F12 data. They are compared with experiment \cite{herndon2005determination}.
All frequencies are given in cm$^{-1}$.}
\end{table}
\clearpage

\subsection{TL: Energy- and force-errors}
\begin{table}[h]
\begin{tabular}{lcccc}
\toprule
& \multicolumn{4}{c}{\bf NN(TL)} \\
& CH$_3$CHO & CH$_3$NO$_2$ & CH$_3$COOH & CH$_3$CONH$_2$\\
\midrule
EMAE:	&	0.0196	&	0.0228	&	0.0027	&	0.0097	\\
ERMSE:	&	0.0197	&	0.023	&	0.0029	&	0.0099	\\
FMAE:	&	0.0119	&	0.0183	&	0.0103	&	0.0161	\\
FRMSE:	&	0.0307	&	0.0366	&	0.0188	&	0.0401	\\
\bottomrule
\end{tabular}
\caption{Out-of-sample errors of the TL models to CCSD(T) quality.
The energy errors are given in kcal/mol and the force errors are given
in kcal/mol/\AA. }\label{sitab:tl_oos_errors}
\end{table}
\clearpage
\subsection{CH$_3$CHO}
\begin{table}[]
\begin{tabular}{rrr}
\toprule
 NN(TL) &  CCSD(T) &  $|\Delta|$ \\
\midrule
       158.34 &   157.97 &       0.37 \\
       503.66 &   503.57 &       0.09 \\
       775.07 &   774.97 &       0.10 \\
       896.09 &   896.02 &       0.07 \\
      1130.36 &  1130.31 &       0.05 \\
      1136.02 &  1136.00 &       0.02 \\
      1388.27 &  1388.30 &       0.03 \\
      1420.92 &  1420.99 &       0.07 \\
      1474.26 &  1474.25 &       0.01 \\
      1484.77 &  1484.90 &       0.13 \\
      1777.53 &  1777.36 &       0.17 \\
      2919.35 &  2919.94 &       0.59 \\
      3030.98 &  3031.38 &       0.40 \\
      3098.85 &  3099.27 &       0.42 \\
      3150.54 &  3151.01 &       0.47 \\
\bottomrule
\end{tabular}
\caption{Normal mode frequencies (in cm$^{-1}$) of CH$_3$CHO calculated from a TL model trained on
CCSD(T) data and compared to their reference \textit{ab initio} values.}\label{sitab:tl_ch3cho_cc_harm}
\end{table}
\clearpage
\subsection{CH$_3$NO$_2$}
\begin{table}[]
\begin{tabular}{rrr}
\toprule
 NN(TL) &  CCSD(T) &  $|\Delta|$ \\
\midrule
        30.33 &    27.39 &       2.94 \\
       474.41 &   474.28 &       0.13 \\
       610.43 &   610.26 &       0.17 \\
       662.28 &   662.25 &       0.03 \\
       929.12 &   929.13 &       0.01 \\
      1114.68 &  1114.60 &       0.08 \\
      1144.23 &  1144.29 &       0.06 \\
      1410.41 &  1410.48 &       0.07 \\
      1425.35 &  1425.33 &       0.02 \\
      1477.67 &  1478.01 &       0.34 \\
      1490.72 &  1491.32 &       0.60 \\
      1613.79 &  1613.72 &       0.07 \\
      3082.80 &  3082.90 &       0.10 \\
      3179.75 &  3179.78 &       0.03 \\
      3207.18 &  3207.13 &       0.05 \\
\bottomrule
\end{tabular}
\caption{Normal mode frequencies (in cm$^{-1}$) of CH$_3$NO$_2$ calculated from a TL model trained on
CCSD(T) data and compared to their reference \textit{ab initio} values.}\label{sitab:tl_ch3no2_cc_harm}
\end{table}
\clearpage
\subsection{CH$_3$COOH}
\begin{table}[]
\begin{tabular}{rrr}
\toprule
 NN(TL) &  CCSD(T) &  $|\Delta|$ \\
\midrule
        80.67 &    80.21 &       0.46 \\
       418.84 &   418.77 &       0.07 \\
       542.51 &   542.63 &       0.12 \\
       582.77 &   582.64 &       0.13 \\
       656.54 &   656.58 &       0.04 \\
       865.99 &   865.99 &       0.00 \\
      1006.75 &  1006.91 &       0.16 \\
      1073.32 &  1073.92 &       0.60 \\
      1216.03 &  1215.93 &       0.10 \\
      1347.96 &  1347.97 &       0.01 \\
      1421.21 &  1421.17 &       0.04 \\
      1484.75 &  1484.64 &       0.11 \\
      1491.69 &  1491.78 &       0.09 \\
      1816.71 &  1816.86 &       0.15 \\
      3059.07 &  3059.07 &       0.00 \\
      3131.34 &  3131.39 &       0.05 \\
      3175.04 &  3175.00 &       0.04 \\
      3756.19 &  3756.31 &       0.12 \\
\bottomrule
\end{tabular}
\caption{Normal mode frequencies (in cm$^{-1}$) of CH$_3$COOH calculated from a TL model trained on
CCSD(T) data and compared to their reference \textit{ab initio} values.}\label{sitab:tl_ch3cooh_cc_harm}
\end{table}
\clearpage
\subsection{CH$_3$CONH$_2$}
\begin{table}[]
\begin{tabular}{rrr}
\toprule
 NN(TL) &  CCSD(T) &  $|\Delta|$ \\
\midrule
        45.28 &    47.22 &       1.94 \\
       250.38 &   240.57 &       9.81 \\
       419.92 &   418.61 &       1.31 \\
       511.15 &   511.53 &       0.38 \\
       550.94 &   551.26 &       0.32 \\
       641.74 &   639.95 &       1.79 \\
       853.02 &   853.02 &       0.00 \\
       985.36 &   984.86 &       0.50 \\
      1058.77 &  1059.14 &       0.37 \\
      1128.27 &  1127.94 &       0.33 \\
      1345.44 &  1345.26 &       0.18 \\
      1411.63 &  1411.86 &       0.23 \\
      1484.27 &  1484.28 &       0.01 \\
      1500.92 &  1500.59 &       0.33 \\
      1630.04 &  1630.02 &       0.02 \\
      1768.18 &  1768.24 &       0.06 \\
      3046.36 &  3045.80 &       0.56 \\
      3119.82 &  3118.47 &       1.35 \\
      3156.39 &  3157.45 &       1.06 \\
      3594.26 &  3595.34 &       1.08 \\
      3730.94 &  3732.00 &       1.06 \\
\bottomrule
\end{tabular}
\caption{Normal mode frequencies (in cm$^{-1}$) of CH$_3$CONH$_2$ calculated from a TL model trained on
CCSD(T) data and compared to their reference \textit{ab initio} values.}\label{sitab:tl_ch3conh2_cc_harm}
\end{table}
\clearpage
\section{Scaled harmonic frequencies}
\begin{figure}[h]
\centering \includegraphics[width=0.9\textwidth]{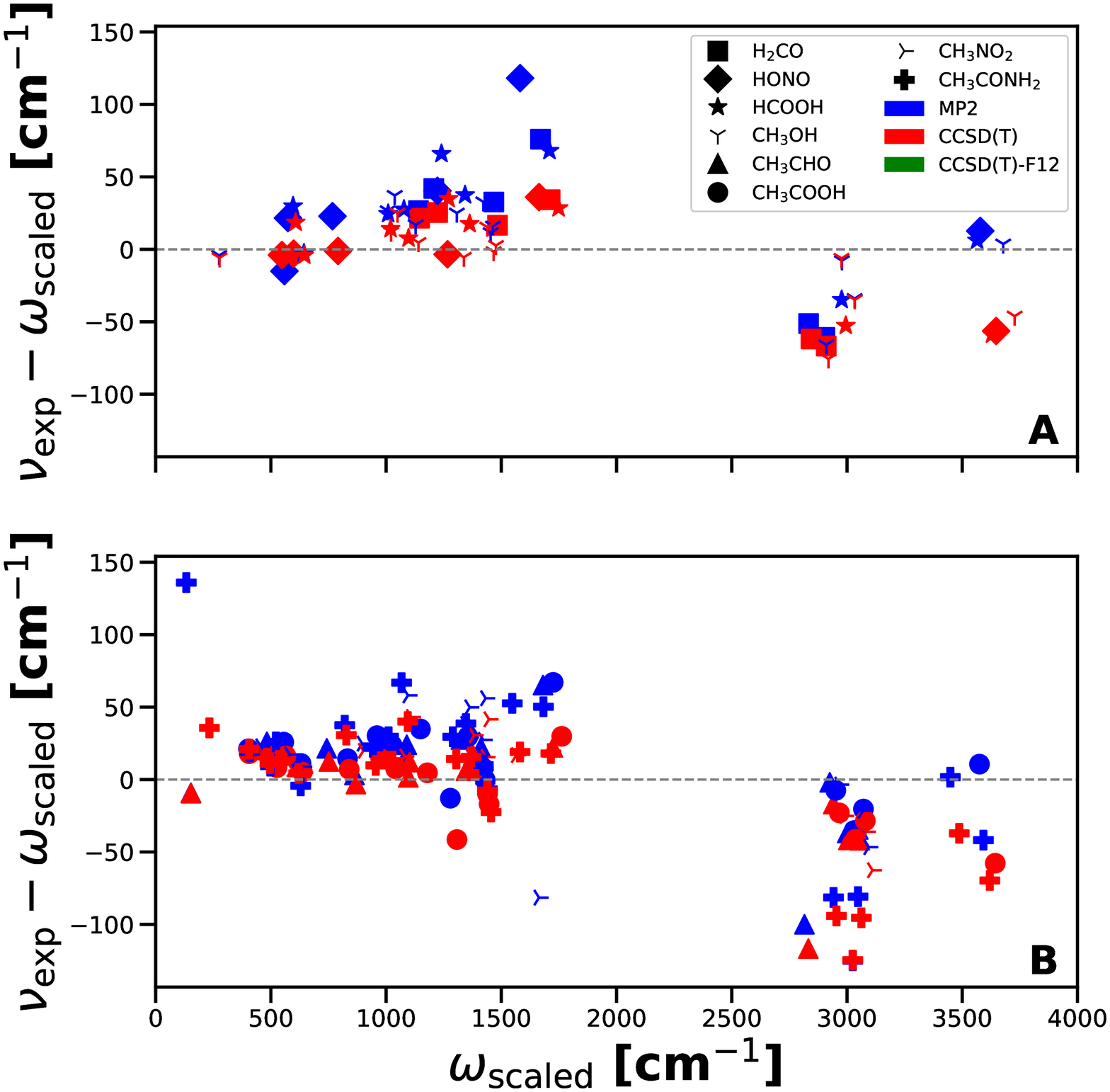}
\caption{The accuracy of the scaled \textit{ab initio} harmonic frequencies is shown with respect to the experimental
values. The figure is divided into two windows for clarity. }
\label{sifig:scaledharm_freq}
\end{figure}
\newpage
\bibliography{references}